\documentclass{article}

\PassOptionsToPackage{square,numbers}{natbib}
\usepackage[preprint]{neurips_2022}

\usepackage[utf8]{inputenc} 
\usepackage[T1]{fontenc}    
\usepackage{hyperref}       
\usepackage{url}            
\usepackage{booktabs}       
\usepackage{amsfonts}       
\usepackage{amsmath}
\usepackage{float}
\usepackage{MnSymbol}%
\usepackage{wasysym}%
\usepackage{nicefrac}       
\usepackage{microtype}      
\usepackage{xcolor}         

\usepackage[english]{babel}
\usepackage{blindtext}

\usepackage[capitalise,noabbrev]{cleveref}
\usepackage{listings}

\usepackage{acro}
\usepackage{relsize}

\usepackage{pifont}
\usepackage{float}
\usepackage{todonotes}

\usepackage[utf8]{inputenc}
\usepackage{pgfplots}
\usepackage[nokeyprefix]{refstyle}

\usepackage{soul}
\usepackage{color}

\DeclareUnicodeCharacter{2212}{−}
\usepgfplotslibrary{groupplots,dateplot}
\usetikzlibrary{patterns,shapes.arrows}
\pgfplotsset{compat=newest} 

\DeclareAcronym{AS}{
    short = AS,
    long = autonomous system,
    short-plural = es ,
    long-plural = s
}

\DeclareAcronym{NAR}{
    short = NAR,
    long = neural algorithmic reasoning,
    short-plural = es ,
    long-plural = s
}

\newcounter{blackcirc}
\newcommand{\blackcircled}[1]{%
\setcounter{blackcirc}{201}\addtocounter{blackcirc}{#1}%
\ding{\theblackcirc}%
}

\usepackage[final]{changes}
\definecolor{changeBlue}{HTML}{D3DEF7}

\title{Learning to Configure Computer Networks with Neural Algorithmic Reasoning}

\author{%
  Luca Beurer-Kellner\textsuperscript{1, \thanks{Correspondence to \texttt{luca.beurer-kellner@inf.ethz.ch}.}}
  \And
  Martin Vechev\textsuperscript{1}
  \And 
  Laurent Vanbever\textsuperscript{1}\\
  \And
  Petar Veličković\textsuperscript{2}\\
  \AND
  \textnormal{\textsuperscript{1}ETH Zurich, Switzerland}\\
  \And
  \textnormal{\textsuperscript{2}DeepMind}\\
  \AND
  \texttt{\url{https://github.com/eth-sri/learning-to-configure-networks}}
}

\begin{document}

\maketitle

\begin{abstract}
  \looseness=-1
  We present a \added{new method} for scaling automatic configuration of computer networks. The key idea is to relax the computationally hard search problem of finding a configuration that satisfies a given specification into an approximate objective amenable to learning-based techniques. Based on this idea, we train a neural algorithmic model which learns to generate configurations likely to (fully or partially) satisfy a given specification under existing routing protocols. By relaxing the rigid satisfaction guarantees, our approach (i) enables greater flexibility: it is protocol-agnostic, enables cross-protocol reasoning, and does not depend on hardcoded rules; and (ii) finds configurations for much larger computer networks than previously possible. Our learned synthesizer is up to $490\times$ faster than state-of-the-art SMT-based methods, while producing configurations which on average satisfy more than 92\% of the provided requirements.  
\end{abstract}

\setlength{\parskip}{4.3pt}

\newcommand{\fwd}{{\scshape Fwd }}
\newcommand{\fwdmat}{\text{{\scshape Fwd}}}

\section{Introduction}
\label{sec:introduction}

\begin{figure}
    \begin{center}
        \includegraphics[width=0.95\textwidth]{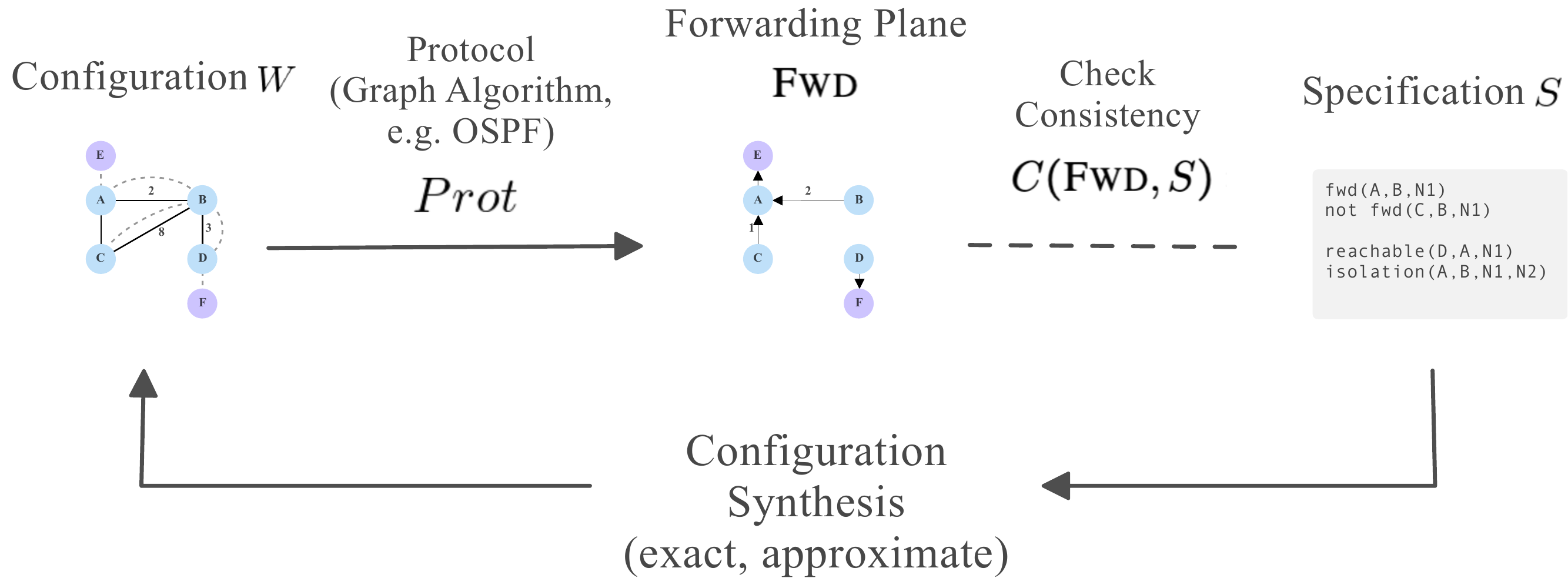}
    \end{center}
    \caption{\textbf{Network configuration synthesis.} The goal of configuration synthesis is to find a configuration $W$ that maximizes consistency $C(\fwdmat,S)$ for a given specification $S$.}
    \label{fig:learning-based-synthesis}
\end{figure}

Configuring large-scale networks is a challenging and important task as network configuration mistakes regularly lead to massive internet-wide outages affecting millions (resp. billions\footnote{As of 2021, Facebook reportedly has 2.9 billion monthly active users~\cite{fbactiveusers2021}.}) of Internet users \cite{fastlyoutage,janardhan2021}. \added{Typically, network operators provide a router-level configuration $W$ which, after applying protocols such as shortest-path routing, induces a certain forwarding behaviour \fwd as illustrated in \cref{fig:learning-based-synthesis}. As this remains a challenging task, much recent research has focused on automating configuration by leveraging synthesis techniques \cite{el2018netcomplete,beckett2017network,narain2008declarative}: A synthesizer is used to automatically generate a router-level configuration $W$ that, after applying routing protocols results in forwarding behavior that satisfies a given specification $S$ on how traffic should be routed.}

\paragraph{SMT-based Synthesis} Due to the hardness of the configuration synthesis problem \cite{bley2007inapproximability}, many effective tools in this domain \cite{el2018netcomplete,el2017network} resort to \added{satisfiability modulo theory (SMT) solvers, which employ search-based procedures to find a solution to a set of first-order logic constraints. This enables comprehensive and exact synthesis by modelling network behavior in first-order logic.} However, these tools are typically protocol-specific, hand-coded, and can exhibit discrepancies in behavior when compared to actual router hardware \cite{birkner2021metha}. Most importantly, however, they can be very slow or fail to complete for large networks. For example, the state-of-the-art SMT-based tool NetComplete \cite{el2018netcomplete} requires more than 6 hours to synthesize a configuration for a network with 64 nodes, for which other SMT-based tools like SyNET \cite{el2017network} take even longer ($>24$ hours) \cite{el2018netcomplete}. Non-SMT-based tools such as Propane~\cite{beckett2016don} or Zeppelin~\cite{subramanian2018synthesis} have achieved better performance, but at the cost of generality. 

\paragraph{\added{Addressing the scalability barrier}} \added{The reason for the wide-spread use of SMT in configuration synthesis is the inherent computational complexity of the underlying synthesis problem, parts of which have been shown to be NP-hard \cite{bley2007inapproximability,fortz2000internet, vissicchio2012ibgp}.} This means that any exact synthesis method is bound to run into scalability issues (as do all SMT-based methods). However, to be practically useful, a synthesizer must scale to the size of real-world networks and the frequency at which configurations are updated. For example, in a Tier-1 ISP, network operators modify their configurations up to 20 times per day, on average \cite{schneider2021snowcap}. We argue that one way to address this scalability barrier, is relaxing the configuration synthesis problem to admit approximate solutions with high utility -- configurations which may not always satisfy all requirements of a given specification but may satisfy almost all of them. Such a configuration would be a much better starting point for a network operator than having no automated support whatsoever. The core technical challenge then is coming up with a strategy likely to find solutions with high~utility.




\paragraph{\added{This Work: enabling fast and scalable configuration} via neural algorithmic reasoning} We address this challenge and present the first learning-based framework for approximate configuration synthesis. Our relaxed formulation allows us, for the first time, to apply end-to-end learning to the problem of network configuration synthesis, \added{thereby enabling fast and scalable configuration synthesis with almost interactive response times (<90s even for large networks).} Technically, we leverage the observation that routing protocols can often be formulated as Bellman-Ford style graph algorithms, a class of problems that has recently been studied in the area of \ac{NAR} \cite{velivckovic2019neural}. Connecting the two fields and building on ideas from \ac{NAR}, we are able to train a graph-based neural model with a strong inductive bias to learn an inverse mapping from specifications back to network configurations: Our model learns how to perform synthesis from a dataset of (specification, configuration) pairs obtained by simulating the involved protocols and observing the computed forwarding state. With this method, we can support cross-protocol reasoning and do not have to manually provide any hardcoded synthesis rules. Concretely, we introduce a generic embedding scheme for topologies and configurations, making our method protocol-agnostic. During synthesis, given a specification, our model predicts distributions of network configurations from which we can sample possible results.


\paragraph{Main Contributions} Our core contributions are:

\begin{itemize}
    \setlength{\itemsep}{4pt}
    \item We formulate a relaxation of the exact configuration synthesis problem, which enables fast and scalable network configuration, amenable to learning-based techniques (\cref{sec:relaxed-config-synthesis}).
    \item We propose a \ac{NAR}-based neural network architecture for learning synthesizer models that rely on a graph-based encoding of topologies and configurations, and a strong inductive bias towards an iterative synthesis procedure (\cref{sec:neural-synthesis-model}).
    \item We conduct an extensive evaluation of our learning-based synthesizer with respect to both precision and scalability. We demonstrate that our learned synthesizer is up to~$490\times$ faster than a state-of-the-art SMT-based tool while producing high utility configurations which on average satisfy $ >93\%$ of provided constraints (\cref{sec:evaluation}).
\end{itemize} 
\section{Configuration Synthesis: Exact and Learned}
\label{sec:relaxed-config-synthesis}

We first state the general configuration synthesis problem and explain why it is hard to solve. We then present a rather different approach based on learning \added{that addresses the scalability barrier of traditional synthesis.}

\paragraph{Forwarding Behavior and Specifications} We focus on the level of the \emph{forwarding plane} of a network. This means we consider how a network forwards traffic, given a packet with a certain destination. The forwarding plane is determined by a distributed computation that depends on the different routing protocols in use. More formally, we define the forwarding plane \fwd as follows:
$$
\text{\fwd} := Prot(W; T)
$$

$Prot(W;T)$ corresponds to the result of applying routing protocols to the network topology $T$ and the configuration $W$ (e.g. link weights). In the following, we omit $T$ as it remains fixed in synthesis. \added{The resulting forwarding plane \fwd can be understood as a directed graph superimposed on topology $T$. It specifies a subset of links that are used to forward packets.
To illustrate consider \cref{fig:learning-based-synthesis}: applying the routing protocols yields forwarding plane \fwd which corresponds to the subset of links $(C,A), (B,A), (A,E)$, $(D,F)$. The other links of the network are not part of the forwarding plane and will thus not be used to forward traffic.}

Given \fwd, we consider a forwarding specification $S := \{R_i\}_i$ as an input to the synthesis problem. Each requirement in $S$ is modelled as a function $R_i$, where $R_i(\fwdmat) = 1$ if \fwd satisfies the requirement and $0$ otherwise. Practical example requirements include reachability, traffic isolation or specifying concrete forwarding paths.

\paragraph{Exact Configuration Synthesis} We formulate the general configuration synthesis problem as the following optimization objective:

\begin{equation}
    \label{eq:config-synthesis}
    W^{\star} := \underset{W \in P(W)}{\arg\max}\; C(Prot(W),S) \quad \text{ where } \quad C(\fwdmat,S) := \frac{\sum_{R_i \in S} \; R_i(\fwdmat)}{|S|}
\end{equation}

$P(W)$ denotes the set of all possible configurations and $C(\fwdmat,S)$ is the \emph{specification consistency} of a forwarding plane $\fwdmat$ w.r.t specification $S$. In traditional, exact configuration synthesis, this objective is solved by limiting the search to globally-optimal configurations such that $C(\fwdmat,S) = 1.0$, i.e. \emph{all} requirements must be satisfied. Such exact methods typically resort to SMT solvers because of the hardness of the underlying problem: configurations comprise a large number of tunable parameters, where the execution of several interacting protocols yields the overall forwarding state. Even worse, parts of the configuration synthesis problem have been shown to be NP-hard \cite{bley2007inapproximability,fortz2000internet, vissicchio2012ibgp}: \added{For example, already the subproblem of finding link weights that yield a given set of forwarding paths under shortest-path routing is NP-hard \cite{bley2007inapproximability}}. This makes scaling exact synthesis to real-world networks extremely challenging.

\paragraph{Learning-Based Synthesis}
\added{To enable fast and scalable configuration synthesis}, we propose to relax both the optimality as well as the rigid satisfaction requirements w.r.t the specification $S$. Concretely, we relax the set of admissible solutions to include configurations that are not optimal, but still satisfy a large number of provided requirements. Note that this is not the same as merely allowing solutions with $C(\fwdmat,S) < 1.0$, because maximum satisfiability does not relax the hardness of the problem. Instead, we propose to search for near-optimal, good solutions and rely on the value of specification consistency $C$ as a measure of quality.

An approximate synthesis formulation relaxes the hardness of the problem, however, it also leads to the difficult technical challenge of finding solutions with high utility (e.g., where many requirements are satisfied). To address this challenge, we propose a \added{rather different} approach where we learn synthesis from data. Concretely, we learn an inverse mapping that attempts to predict approximate solutions $\hat{W}^{\star}$ with high utility, as guided by the following objective:

$$
\hat{W}^{\star} = Prot^{-1}(\fwdmat) \; \text{ s.t. } \; C(\fwdmat,S) \text{ is high}
$$

This can be implemented as a synthesizer model $M_{Syn}$ which produces a solution given just the topology $T$ and the specification $S$:

$$
\hat{W}^{\star} = M_{Syn}(S;T)
$$

We propose a learning-based approach for training such synthesizer models based on neural algorithmic reasoning (cf. \cref{sec:neural-synthesis-model}). First, however, we discuss the routing protocols that make up function $Prot$ and how they relate to graph algorithms and by extension to \ac{NAR}.

\begin{figure*}
    \centering
    \includegraphics[width=0.8\textwidth]{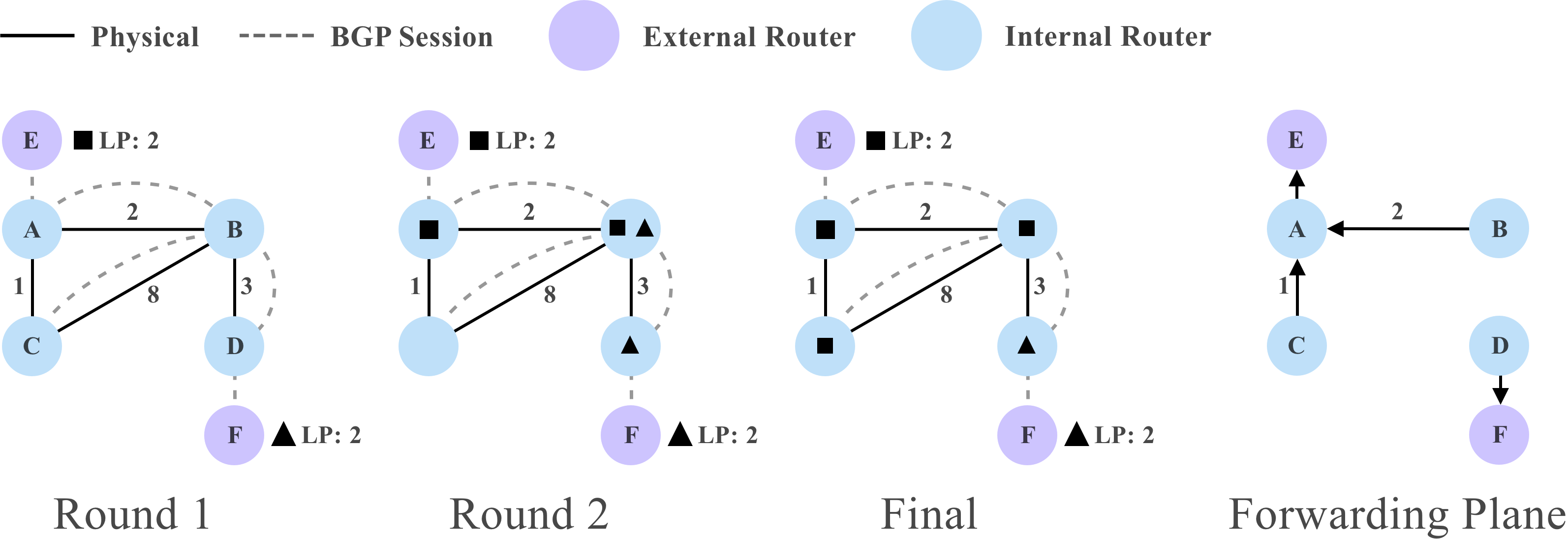}
    \caption{An illustration of propagating BGP route announcements in a small network, including internal peers (blue), external peers (purple), BGP sessions as well as physical links. The last graph illustrates the resulting forwarding plane after applying BGP/OSPF.}
    \label{fig:network-fwd-tree}
\end{figure*}

\section{Routing Protocols as Graph Algorithms}
\label{sec:bgp-ospf}
\label{sec:synthesis-setting}

\newcommand{\as}{AS~}
\newcommand{\ases}{ASes~}

\newcommand{\squareann}{{\tiny$\blacksquare\;$}}
\newcommand{\triangleann}{{\small$\blacktriangle\;$}}

In practice, the routing protocols defining $Prot$ are implemented as distributed systems in which multiple peers communicate to determine the network's forwarding state. However, theoretical work on routing algebras \cite{griffin2005metarouting} has shown that the underlying computation can be understood as a traditional message-passing graph algorithm. As a consequence, many routing protocols can be formulated as Bellman-Ford (BF) style propagation processes. This class of problems has also recently been subject to work on \ac{NAR} \cite{velivckovic2019neural,velivckovic2021neural} and algorithmic alignment \cite{xu2019can}. \added{The authors of these works demonstrate that neural networks are capable of closely imitating BF-style algorithms when provided with a suitable inductive bias. Based on this insight, NAR proposes to replace traditional algorithms with neural networks to learn improved algorithmic procedures or extend existing algorithms to be applicable to raw data \cite{velivckovic2021neural}. Following the idea of \ac{NAR}, we implement a synthesizer model for network configurations as an iterative Graph Neural Network (GNN) to learn $Prot^{-1}$ by relying on a Bellman-Ford style inductive~bias.}

\paragraph{Synthesis Setting} Our approach is general for the domain of networks, but we focus on two widely-used routing protocols: \added{(1) Open Shortest Path First (OSPF) \cite{moy1994ospf} -- it uses link weights to route traffic along the shortest path towards the destination, and the (2) Border Gateway Protocol (BGP)} \cite{rekhter1994border}, used to exchange reachability information, mostly on the level of larger backbone networks. When using BGP, a routing destination announces its existence to other networks and routers by sending out BGP announcements. Receivers of announcements then choose to pass them on to other peers, redistribute them internally and/or modify them according to a set of decision rules. BGP and OSPF interact, for instance, BGP will consider the OSPF cost of internal destinations in its routing decisions. \added{This means, that effective BGP/OSPF synthesis tools must implement cross-protocol reasoning, configuring BGP and OSPF, such that together they yield the desired forwarding behaviour.} 




\paragraph{Example}
\added{
To provide a basic intuition about BGP/OSPF routing, consider the simple example network in \cref{fig:network-fwd-tree}. We apply the two protocols to obtain the forwarding plane. Physical OSPF edges are labelled with a corresponding link weight determining the OSPF cost of paths through the network. Dotted lines represent designated BGP edges, used to propagate BGP announcements. Our network imports multiple BGP announcements \squareann and \triangleann from E and F respectively. Both represent a route to the same destination. As shown, the announcements have a so-called BGP local preference of 2 (we ignore other BGP properties in this example). The announcements are propagated through the network and the best route is selected according to a designated BGP decision procedure. \cref{fig:network-fwd-tree} shows the intermediate states of this propagation process. As both announcements have the same local preference value, the decision cannot eliminate based on that. Instead, in round 2, node B selects \squareann over \triangleann due to a lower OSPF cost (shorter path) of 2 via node A as compared to 3 via node D. Node D selects \triangleann over \squareann, since it learns this route directly from an external peer which is preferred in BGP. After Round 3, the BGP propagation process converges to a stable state and we can derive the forwarding plane as shown on the left in \cref{fig:network-fwd-tree}. For completeness, we include the full BGP decision process in \cref{app:bgp-decision-process}.
}

\paragraph{Configuration Parameters} For our purposes, we define the set of synthesized configuration parameters as follows: For OSPF, we synthesize link weights as explored in existing work \cite{el2018netcomplete,fortz2000internet}. For BGP, we focus on a setting, where we synthesize BGP import policies only. This means we synthesize the modifications required for BGP announcements when entering the network, to satisfy the routing specification. Previous work has confirmed that this is a realistic configuration setting that applies to a majority of real-world networks \cite{caesar2005bgp,cittadini2010doing,  steffen2020probabilistic}.



\looseness=-1 Based on the observation that routing protocols such as OSPF and BGP can be expressed as \added{message-passing algorithms}, we heavily rely on GNNs/\ac{NAR} for the design of our synthesizer model as discussed next. In our evaluation, we then train such a model for the concrete case of BGP/OSPF and compare synthesis performance with a traditional SMT-based tool.


\section{Neural Configuration Synthesis Model}
\label{sec:neural-synthesis-model}

\newcommand{\comp}[1]{\text{{\scshape #1}}}

\begin{figure}
    \centering
    \includegraphics[width=1.0\textwidth]{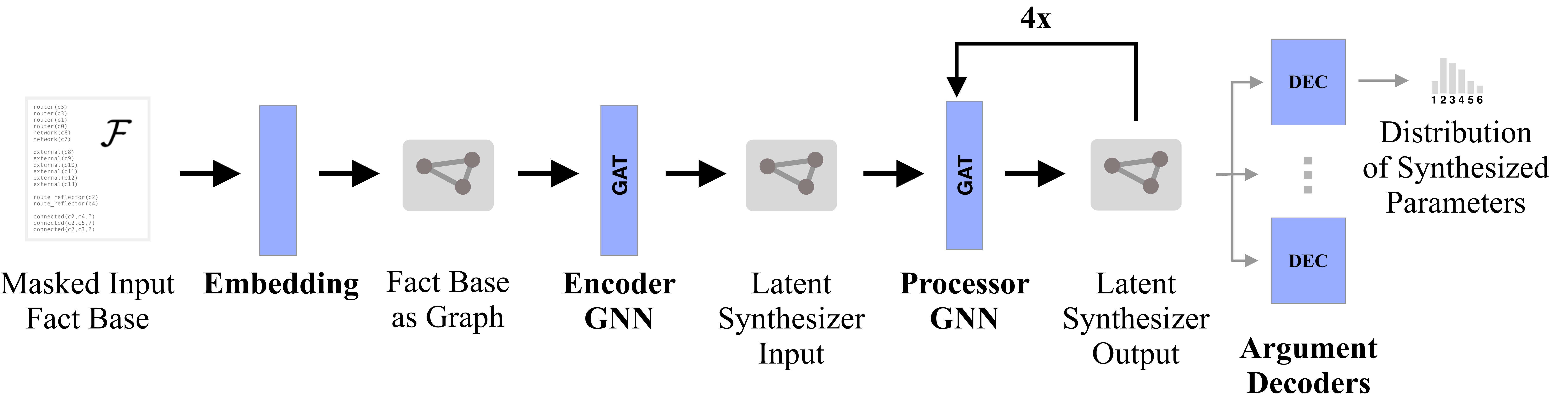}
    \caption{\textbf{Neural synthesizer architecture}. A provided input fact base is first embedded as a graph including the masked to-be-synthesized parameters. Then encoder GNN, processor GNN and decoder networks are applied to obtain a distribution over synthesized parameters.}
    \label{fig:architecture}
\end{figure}

We train a graph neural network (GNN) as the neural synthesis configuration model. The model's input consists of a topology, a specification, and a configuration sketch where to-be-synthesized parameters are omitted. Following the NAR paradigm, we first encode this synthesizer input in latent space. Then, we apply an iterative processor network based on the graph attention mechanism \cite{velivckovic2017graph}. Last, we apply a decoder network to predict the values of omitted configuration parameters, thereby synthesizing a configuration. To remain protocol-agnostic, our model is generic with respect to the input format, using an intermediate representation based on Datalog-like facts. \cref{fig:architecture} provides an overview of our graph-based neural synthesizer architecture.

\subsection{Training Dataset of Inverse Pairs}
\label{sec:dataset-model}

A learning-based synthesizer model is formulated as a \added{supervised learning problem}. Thus, we can directly train a neural network to learn the inverse mapping, given a dataset of corresponding input-output pairs. To obtain such a dataset, we sample a random network configuration for some topology using a uniform generative process. Then, we simulate the involved protocols using $Prot$, to obtain the corresponding forwarding plane. Next, we extract a specification by randomly selecting properties that hold for the computed forwarding plane. This leaves us with a pair of specification and topology as input, and a corresponding configuration as output.

The key to constructing the dataset is the implementation of $Prot$. Even though $Prot$ is protocol-specific, it turns out the overall implementation effort is comparatively low, especially when compared to SMT-based synthesis methods. Protocols are well-defined algorithms that can be easily simulated, whereas the alternative of implementing hardcoded synthesis rules directly often requires expert knowledge of SMT solvers. This process may also be adapted to rely on actual router hardware to compute the result of $Prot$, thereby capturing real-world behavior precisely.


\subsection{Embedding Topologies, Specifications and Configurations}
\label{sec:graph-embedding}
To remain agnostic with respect to routing protocols, our model architecture implements a generic graph-based encoding of Datalog-like facts, similar to knowledge graphs \cite{rosso2020beyond}: we first encode topologies, configurations and specifications as a set of Datalog-like facts and then employ a generic embedding scheme to embed these facts into latent space.

\begin{figure}
    \centering
    \begin{minipage}[t]{0.3\textwidth}
    \begin{lstlisting}[basicstyle=\ttfamily\small, escapeinside={(*}{*)}]
router(A)
router(B)
(...)

conn(A,B,2)
conn(A,C,?)(*\blackcircled{1}*)
(...)
    \end{lstlisting}
\end{minipage}
\begin{minipage}[t]{0.35\textwidth}
    \begin{lstlisting}[basicstyle=\ttfamily\small, escapeinside={(*}{*)}]
network(N1)
bgp_route(E,N1,2,3,1,0,1)

fwd(A,B,N1)
not fwd(B,A,N1)
(...)
    \end{lstlisting}
\end{minipage}
    \caption{A fact base encoding a network's topology, parameters such as link weights (\texttt{conn} facts) and a specification (\texttt{fwd} facts).}
    \label{fig:datalog-encoding}
\end{figure}

\paragraph{Fact Base} A set of Datalog-like facts, as depicted in \cref{fig:datalog-encoding}, serves as the input \emph{fact base} $\mathcal{F}$ to our synthesizer model. $\mathcal{F}$ is a set of facts $f(a_0, \dots, a_n)$ where arguments may be constants (e.g. A or B) or integer literals. Each fact has a corresponding boolean truth value denoted as $[f(a)]_{\mathcal{B}}$. For instance, from the fact base in \cref{fig:datalog-encoding} we can derive $[fwd(A,B,N1)]_{\mathcal{B}} = \text{\texttt{true}}$ and $[fwd(B,A,N1)]_{\mathcal{B}} = \text{\texttt{false}}$.

\paragraph{Synthesis as Completion Task} A network's topology, protocol configuration, parameters, and the specification are all represented in a single fact base. To predict the value of unknown, to-be-synthesized parameters, we support the notion of unknown parameters as illustrated at \blackcircled{1} in \cref{fig:datalog-encoding}, where the link weight between router nodes \texttt{A} and \texttt{C} is omitted. Based on this input format, synthesis corresponds to using our model to predict the value of unknown parameters in a provided fact base.

\begin{figure}

    \definecolor{topologyBlue}{rgb}{0.69,0.737,1.0}
    \definecolor{specGreen}{HTML}{CBF8CA}
    \definecolor{configYellow}{HTML}{FFF6B2}
    \DeclareRobustCommand{\hltopo}[1]{{\sethlcolor{topologyBlue}\hl{#1}}}
    \DeclareRobustCommand{\hlspec}[1]{{\sethlcolor{specGreen}\hl{#1}}}
    \DeclareRobustCommand{\hlconfig}[1]{{\sethlcolor{configYellow}\hl{#1}}}

    \begin{minipage}[t]{1.0\textwidth}
        \begin{minipage}{0.5\textwidth}
            \centering
            \includegraphics[width=0.9\textwidth]{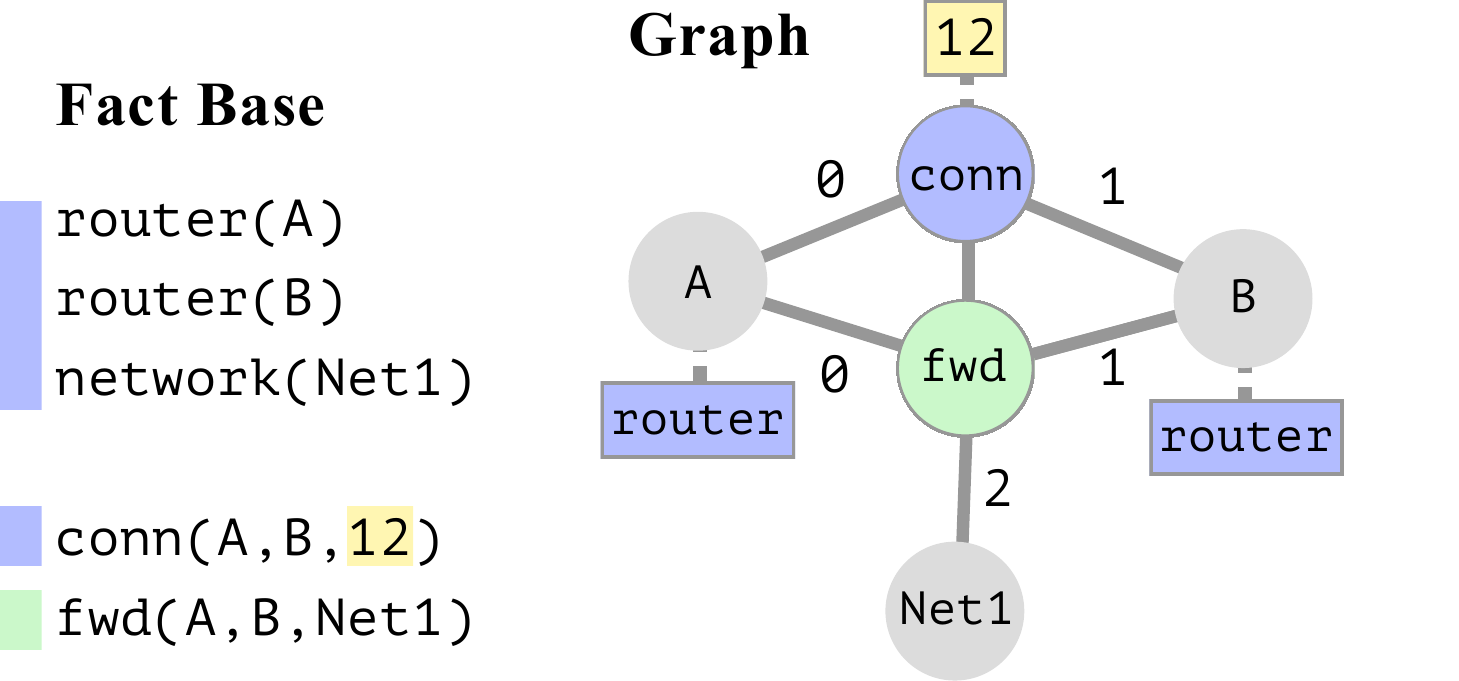}
        \end{minipage}
        \begin{minipage}{0.4\textwidth}
        \footnotesize 
        \textbf{Node Features}\vspace{0.5cm}\\
        $\begin{array}{rcl}
                    h_A &:=& emb(router) = \mathbf{V_{\text{router}}}\\
                    h_B &:=& emb(router) = \mathbf{V_{\text{router}}}\\
                    h_{\text{Net1}} &:=& emb(network) = \mathbf{V_{\text{network}}}\\
                    h_{\text{conn(A,B,12)}} &:=& emb(conn) + emb(conn, 2, 12)\\
                    &:=& \mathbf{V_{\text{conn}}} + \mathbf{W_{conn_2}} onehot(12) \\
                    h_{\text{fwd(A,B,Net1)}} &:=& emb(fwd) = \mathbf{V_{\text{fwd}}}\\
        \end{array}$
        \end{minipage}
    \end{minipage}
    
    \caption{An example of embedding a simple fact base using our generic graph embedding scheme. Different neighborhoods are indicated as edge labels \texttt{0,1,2}. The full set of structural embedding rules is provided in \cref{fig:graph-embedding}, \cref{app:embedding-scheme}. We mark nodes and facts relating to \hltopo{topology}, \hlspec{specification} and \hlconfig{configurations} in color.}
    \label{fig:graph-encoding-example}
\end{figure}

\paragraph{Embedding} To transform a fact base into a graph with node features, we apply a structurally-defined embedding scheme. \added{An example of this embedding scheme is given in \cref{fig:graph-encoding-example}. As shown, we embed topology, specification and configuration all into a single graph. This places network routers, the related specification predicates as well as configuration parameters in close adjacency to each other, simplifying the synthesis procedure for the processor network. In our scheme, both} facts and constants are represented as distinct nodes. The relationships between facts and constants are encoded as node adjacency as defined by neighborhood functions $N_i$. We use multiple neighborhoods, i.e. multiple types of edges, to encode the argument position of a constant occurring in a fact. To handle unknown parameters, we replace the corresponding embedding with a learned $\mathbf{V_{hole}}$ embedding. Overall, the embedding function $\comp{Emb}$ relies on a set of learned parameters, including $\mathbf{V_f} \in \mathbb{R}^D$ per fact type in~$\mathcal{F}$, $\mathbf{W_{\text{bool}}} \in \mathbb{R}^{D \times 2}$ for boolean values, $\mathbf{W_{f_i}} \in \mathbb{R}^{D \times N}$ per integer argument of fact type $f$ and $\mathbf{V_{hole}} \in \mathbb{R}^D$ to represent unknown parameters. $D$ is the dimensionality of the latent space and $N$ specifies the number of supported integer values. For the complete embedding scheme, please see \cref{app:embedding-scheme}.



\subsection{A NAR-based Synthesizer Model}
\label{sec:graph-transformer}


Our synthesizer model employs an encode-process-decode architecture, inspired by neural algorithmic reasoning \cite{velivckovic2021neural}. It employs four main components to produce the output distribution for an unknown parameter in a fact base $\mathcal{F}$: The fact base embedding $\comp{Emb}$, the encoder $\comp{Enc}_{GAT}$, the processor $\comp{Proc}_{GAT}$ and the fact-type specific decoder network $\comp{Dec}_{f_i}$.

Overall, the model can be expressed as follows, where $X_j$ and $H_j$ refer to the intermediate node representations per node $j$ in the graph constructed via our fact base embedding.

\begin{equation}
    \begin{aligned}
        X_j &:= \comp{Enc}_{GAT}(\comp{Emb}(\mathcal{F}))_j + \textbf{z}& \text{where }\;\; \textbf{z} \sim \mathcal{N}(0,1) &\\
        H_j &:= \comp{Proc}_{GAT}(\{X_i\})_j\\
        O_{f_j(a_0,\dots,a_{i-1},?,\dots,a_n)} &:= softmax(\comp{Dec}^f_{i}(H_j))
    \end{aligned}
\end{equation}

$\comp{Enc}_{GAT}$ is a GNN relying on Graph Attention Layers (GAT) \cite{velivckovic2017graph} for propagation. We additionally apply noise to the latent node representation $X_j$ via $\mathbf{z}$ as a source of non-determinism, anticipating the fact that the synthesis problem often has more than one solution. The processor $\comp{Proc}_{GAT}$ is modelled as an iterative process. It consists of a 6-layer graph attention module which we apply for a total of 4 iterations. According to the NAR paradigm, this computational structure encodes an inductive bias towards an iterative solution to the synthesis problem.

In the encoder and processor GNNs, we rely on a composed variant of the graph attention layer as introduced by \cite{velivckovic2017graph}. We employ multiple graph attention layers in lockstep, one per neighborhood $N_i$ defined by our graph embedding (cf. \cref{sec:graph-embedding}), and combine the intermediate results after each step by summation. For details on the graph attention module, please refer to \cref{app:gat-model-detail}.

Finally, an argument decoder $\comp{Dec}_{f_i}$ is used, to produce the output distribution $O_{f_j(a_0,\dots,a_{i-1},?,\dots,a_n)}$ for an unknown parameter at position $i$ of fact $f_j \in \mathcal{F}$, corresponding to node $j$. For this, the synthesizer model provides one decoder per integer argument, per supported fact type. For example, a model for the fact base in \cref{fig:datalog-encoding} would provide a decoder network $D^{\text{conn}}_2: \mathbb{R}^D \rightarrow \mathbb{R}^N$ to decode values for the third argument of a \texttt{conn} fact. Decoder networks are implemented as simple multi-layer perceptron models.

\paragraph{Supervision Signal} During training, we mask all fact arguments that represent to-be-synthesized configuration parameters in our synthesis setting as unknown parameters. As a supervision signal we use the masked values as the ground-truth and apply a negative log-likelihood loss to the output distribution of the corresponding argument decoder networks.




\paragraph{Multi-Shot Sampling}
\label{sec:decoding-sampling}

To sample values from the output distributions of unknown parameters, we apply a multi-shot sampling strategy. We sample values only for a subset of unknown parameters, insert them in the input fact base and run the synthesizer again. We repeat until no more unknown parameters remain. This multi-shot strategy allows the model to incorporate concrete values of previously synthesized parameters in its computation. In our evaluation, we compare this approach to sampling all parameters at once.


\begin{table}
    \caption{Average consistency with $3\times 16$ BGP/OSPF requirements, sampling configurations randomly from a uniform distribution and multi-shot sampling using our synthesizer model.}
    \centering
    \footnotesize
       \begin{tabular}{rrcccccccccc}
        \toprule
       & Random & 1-Shot & 4-shot & 8-shot \\
       \toprule
       Small & 0.87$\pm$0.11 & 0.94$\pm$0.04 & \textbf{0.95}$\pm$0.03 & \textbf{0.95}$\pm$0.04 \\
       Medium & 0.81$\pm$0.10 & \textbf{0.96}$\pm$0.04 & \textbf{0.96}$\pm$0.04 & \textbf{0.96}$\pm$0.04 \\
       Large & 0.80$\pm$0.05 & 0.93$\pm$0.05 & 0.93$\pm$0.05 & \textbf{0.94}$\pm$0.05 \\
   \end{tabular}
   \label{fig:eval-consistency-baseline}
\end{table}

\section{Evaluation}
\label{sec:evaluation}

In this section, we assess the performance of our model trained for BGP/OSPF synthesis. For this, we trained a synthesizer model on a dataset of $10,240$ samples, constructed by randomly sampling topologies, corresponding specifications and BGP/OSPF configurations as described in \cref{sec:dataset-model}. For details on BGP/OSPF dataset generation and training, please see \cref{app:training}. \added{Lastly, we also evaluate our architectural design decisions in an ablation and parameter study in \cref{app:additional-experiments}}.


\paragraph{Dataset, Metrics and Experimental Setup} We compare using datasets Small (S), Medium (M), and Large (L). Each dataset comprises 8 real-world topologies taken from the Topology Zoo \cite{knight2011internet}, where the number of nodes lies between 0-18, 18-39, and 39-153, respectively. To obtain random forwarding specifications we use the same generative pipeline as discussed in \cref{sec:dataset-model}. Regarding forwarding requirements, we implement support for three specification facts: \texttt{fwd} requirements to set/block forwarding paths, \texttt{reachable} to specify reachability and \texttt{trafficIsolation} to induce traffic isolation among traffic classes (no shared links). In each topology, we do synthesis for 4 different traffic classes (routing destinations) at a time. \added{Overall, this results in 3 datasets x 8 topologies per dataset x 3 differently-size specifications = 72 synthesis tasks}. To assess synthesis quality, we determine specification consistency as the relative number of specification facts in a fact base that are satisfied by the synthesized configuration. We run all experiments on an Intel(R)~i9-9900X@3.5GHz machine with 64GB of system memory and an NVIDIA RTX 3080 GPU with 10GB of video~memory.

\subsection{Synthesis Quality}
\label{sec:synthesis-quality}



\begin{table}
    \caption{Average specification consistency of our synthesizer model, where standard deviation is reported with respect to the different topologies in a dataset. We apply the model to synthesis tasks with $3\times N$ requirements, i.e. $N$ requirements per supported specification fact.}
   \label{fig:eval-consistency}
    \begin{center}
       \footnotesize
       \begin{tabular}{rrllllllllll}
       \toprule
       \multicolumn{2}{l}{Dataset} & \tiny\texttt{fwd} & \tiny\texttt{reachable}  & \tiny\texttt{trafficIsolation} & Overall & Full Matches & $>90\%$ Matches\\
       \midrule
       $3\times 2$ & S & 0.97 & 0.94 & 1.00 & \textbf{0.96}$\pm$0.07 & 6/8 & 6/8\\
& M & 0.95 & 0.94 & 1.00 & \textbf{0.94}$\pm$0.08 & 5/8 & 5/8\\
& L & 0.92 & 1.00 & 1.00 & \textbf{0.94}$\pm$0.06 & 4/8 & 4/8\\
\midrule
$3\times 8$ & S & 0.98 & 0.98 & 0.91 & \textbf{0.96}$\pm$0.05 & 4/8 & 7/8\\
& M & 0.97 & 0.98 & 1.00 & \textbf{0.98}$\pm$0.03 & 4/8 & 8/8\\
& L & 0.96 & 0.92 & 0.97 & \textbf{0.95}$\pm$0.03 & 1/8 & 8/8\\
\midrule
$3\times 16$ & S & 0.98 & 0.92 & 0.95 & \textbf{0.95}$\pm$0.03 & 2/8 & 8/8\\
& M & 0.95 & 0.95 & 0.98 & \textbf{0.96}$\pm$0.04 & 3/8 & 7/8\\
& L & 0.94 & 0.91 & 0.95 & \textbf{0.93}$\pm$0.05 & 1/8 & 6/8\\
   \end{tabular}\end{center}
\end{table}

To assess the quality of synthesized network configurations, we examine specification consistency with increasingly large topologies and specifications. For each synthesis task, we run our synthesizer 5 times using 4-shot sampling and report the network configurations with the highest overall consistency. \cref{fig:eval-consistency} documents the results. On average, our synthesis model achieves $>93\%$ specification consistency for all datasets and specifications. With few requirements, the synthesizer model even succeeds in producing fully-consistent configurations (cf. Full Matches in \cref{fig:eval-consistency}). We observe a slight decrease in consistency with increasingly large topologies.


\paragraph{Multi-Shot Sampling} To determine the effect of multi-shot-sampling, we report consistency when using 1-shot, 4-shot and 8-shot sampling in \cref{fig:eval-consistency-baseline}. As a baseline, we also show consistency when sampling configurations from a uniform distribution (per parameter). \added{This simple method can achieve suprisingly good results, as parts of a specification may be satisfied naturally by the mechanics of shortest-path routing. Still, competing with this baseline our synthesizer model shows clear improvements and multi-shot sampling further increases consistency.}

\paragraph{Number of Samples} We also consider the number of times we sample from our synthesizer model. We observe that sampling more than one alternative configuration can lead to improved best-of specification consistency across all datasets. \added{Based on this observation, we boost synthesis peformance by sampling multiple times per synthesis task. For each configuration that we obtain in this way, we simulate the routing protocols, obtain the forwarding plane and check specification consistency. This fully automated process allows us to determine the best result without consulting the user. We select the best result as the overall output for synthesis, dismissing the other samples.} We experiment how the number of times we sample affects the resulting consistency in \cref{app:more-eval}. Overall, sampling more than once is beneficial for all datasets. Average best consistency values appear to be reached after 4-5 samples for $3\times 16$ BGP/OSPF requirements. Hence, \added{we rely on 5 samples in all other experiments as a trade-off of fast synthesis time and good specification consistency.}

\paragraph{Unsatifiable Specifications} In practice, network operators may sometimes provide unsatisfiable specifications. While exact methods will typically return an error for such inputs, our relaxed setting enables us to consider partial solutions, i.e. configurations that still achieve high specification consistency while ignoring unsat requirements. This may be preferable in some scenarios, especially when perfect specification consistency is not critical anyway. To simulate this scenario, we evaluate specification consistency for OSPF-only synthesis tasks, which were all verified to be unsatisfiable using the SMT-based synthesizer NetComplete \cite{el2018netcomplete}. We still observe a comparatively high average consistency of $0.90$ for our learned synthesizer. This suggests that our model is indeed capable of handling unsatisfiable specifications, while still producing good, partial solutions. For more detailed unsat results and methodology, see \cref{app:eval-unsat}.

\begin{table}
    \caption{Comparing consistency and synthesis time of our method (Neural) with the SMT-based NetComplete. The notation {\tiny{n/8 TO}} indicates the number of timed out runs out of 8 (25+ minutes).}
    \label{fig:evaluation-compare-bgp}
    \footnotesize
    \begin{center}
        \footnotesize
        \begin{tabular}{rrlllllllll}
        \toprule
        \multicolumn{2}{l}{\# Requirements} & NetComplete (s) & Neural CPU (s) & Speedup &$\emptyset$ Consistency& Full Matches\\
        \midrule
        2 reqs. & S & 18.07$s\pm$14.55 & 0.72s$\pm$0.54 & \textbf{25.2x} & 0.97$\pm$0.09 & 7/8\\
         & M & 60.86$s\pm$33.39 & 3.18s$\pm$4.32 & \textbf{19.1x} & 0.94$\pm$0.13 & 6/8\\
         & L & 1389.48$s\pm$312.58\tiny{  7/8 TO} & 24.25s$\pm$28.35 & \textbf{57.3x} & 0.99$\pm$0.03 & 7/8\\
        \midrule
        8 reqs. & S & 247.69$s\pm$436.90 & 1.25s$\pm$1.02 & \textbf{198.7x} & 0.96$\pm$0.08 & 6/8\\
         & M & >25m\tiny{  8/8 TO} & 4.55s$\pm$4.30 & \textbf{329.8x} & 0.97$\pm$0.04 & 4/8\\
         & L & >25m\tiny{  8/8 TO} & 31.28s$\pm$28.53 & \textbf{48.0x} & 0.97$\pm$0.05 & 5/8\\
        \midrule
        16 reqs. & S & 1416.83$s\pm$235.25\tiny{  7/8 TO} & 2.88s$\pm$1.66 & \textbf{492.0x} & 0.92$\pm$0.06 & 1/8\\
         & M & >25m\tiny{  8/8 TO} & 6.53s$\pm$5.10 & \textbf{229.8x} & 0.95$\pm$0.05 & 2/8\\
         & L & >25m\tiny{  8/8 TO} & 87.99s$\pm$141.97 & \textbf{17.0x} & 0.95$\pm$0.03 & 2/8\\
    \end{tabular}\end{center}
\end{table} 

\subsection{Comparison to SMT-based Synthesis}
\label{sec:synthesis-vs-smt}

We compare the synthesis time of our learned synthesizer with the SMT-based, state-of-art configuration synthesis tool NetComplete \cite{el2018netcomplete}. We compare BGP/OSPF and OSPF-only synthesis time. For each topology/specification, we report the total time of running our synthesizer 4 times using 4-shot sampling, whereas for consistency we report the best out of the 4 runs. If a fully consistent configuration is found, we do not continue to sample and only report time until then. For NetComplete, we time out at 25 minutes and report that as a lower bound if exceeded.
We restrict our comparison to specifications that only include forwarding paths (i.e. \texttt{fwd} facts), as the type of supported requirements in NetComplete does not fully align with our model.
%
%

\cref{fig:evaluation-compare-bgp} shows that our synthesizer model outperforms NetComplete by multiple orders of magnitude. We observe a speedup of $20-490\times$ that increases with the size of topology/specification. \added{The loss of precision due to approximation remains moderate, with the average consistency of the synthesized configurations being greater than $92\%$ even for large topologies.}
In comparison, NetComplete times out on more than half of the synthesis tasks, which means that we only observe a lower bound for speedup. Running our model on a GPU can result in even greater speedups (up to $900\times$ given enough GPU memory, cf. \cref{app:bgp-gpu-performance}). Comparing OSPF-only synthesis time, our synthesizer model achieves a speedup of 2-10x over NetComplete. With 16 requirements on dataset L this can also increase up to $500\times$. Further, our model produces fully-consistent OSPF configurations even more often than for BGP/OSPF. See \cref{app:evaluation-compare-ospf} for the full OSPF-only synthesis comparison.

\subsection{Discussion} We have shown that our learning-based synthesizer reaches a very high degree of consistency, often producing fully-consistent configurations. With respect to synthesis time, we outperform SMT-based methods by a large margin, especially for larger topologies. However, this comes at a price: our model sometimes fails to produce fully consistent configurations, especially for large topologies with many requirements. In comparison, SMT-based synthesis will always produce fully consistent configurations if and once it completes. We, therefore, observe a trade-off between consistency and synthesis time. We also note that our evaluation considers real topologies (Topology Zoo \cite{knight2011internet}) but not real specifications. This is due to the lack of a large, practical dataset thereof. Still, we experiment with robustness (\cref{app:additional-experiments}) by introducing distribution shift regarding the size of specifications during training and achieve comparable performance. 

\paragraph{Scaling to Even Larger Topologies} The Topology Zoo \cite{knight2011internet} as used for our evaluation, provides a good range of realistically-sized networks. However, if we consider synthesis at very large scale (e.g. thousands of routers), we note the following scalability limitations: (1) Our models consume a lot of video memory, reaching beyond the amounts available on current consumer GPUs ($>12$GB). This limit is reached with networks of $150$ or more routers, and we have to do inference on the CPU (as indicated in \cref{fig:evaluation-compare-bgp}). If even faster synthesis is important at this scale, more than one such GPU is needed for inference. Further, (2) the distance that information is propagated in the synthesizer GNN is finite due to the model's fixed number of iterations. This means that in very large networks, our synthesizer model will only be capable of deriving solutions by local reasoning which is very likely to impact synthesis quality. While out of the scope of this paper, longer training with more synthesizer iterations and larger topologies may be necessary to obtain comparable results at very large scale.

Nonetheless, we envision a wide range of practically-relevant applications for fast, approximate synthesis, including ML-guided synthesis, unsatisfiable specifications, and hybrid synthesizers, leveraging both learning and SMT solvers. We list a number of future directions in \cref{app:future}.

\section{Related Work}
\label{sec:related-work}

\paragraph{Traditional Configuration Synthesis} 
Next to methods based on compilation such as Propane/PropaneAT \cite{beckett2017network}, there is a number of exact configuration synthesis methods based on constraint solving, e.g. ConfigAssure \cite{narain2008declarative}, SyNet \cite{el2017network} and NetComplete \cite{el2018netcomplete}. These tools are typically hand-coded, very protocol-specific, and can be slow due to the solvers they employ.
%
In contrast, our approach is approximate but scales to much larger networks. Further, as a side-product of our learning-based approach, we can easily adapt to new protocols and allow for transparent cross-protocol reasoning, merely by training on different protocol data.


\paragraph{GNNs and Networking} DeepBGP \cite{bahnasy2020deepbgp} relies on GNNs and reinforcement learning to do configuration synthesis. However, it is limited to BGP configuration and is slower than SMT-based synthesis. In contrast, our learning-based framework is cross-protocol, does not rely on reinforcement learning and provides better synthesis times. Apart from configuration synthesis, GNNs have also been applied to other problems in the networking domain. For example, the authors of RouteNet \cite{rusek2020routenet} use GNNs to predict networking performance metrics. Other work focuses on learning improved protocols like Q-Routing \cite{boyan1994packet} and Graph-Query Neural Networks \cite{geyer2018learning}.

\paragraph{Neural Algorithmic Reasoning} 
\ac{NAR} \cite{velivckovic2021neural} refers to the idea of replacing algorithms with neural networks to learn improved algorithmic procedures. Successful applications include graph algorithms \cite{velivckovic2019neural}, combinatorial optimization problems \cite{cappart2021combinatorial,joshi2020learning,selsam2019guiding} and multi-task settings \cite{ibarz2022}. Our synthesis framework is the first application of NAR to the networking domain and relies on the NAR-native encode-process-decode architecture. In the hierarchy of NAR approaches in \cite{cappart2021combinatorial}, our method is an algorithm-level approach, as we do not supervise on intermediate steps. Although step-level methods promise better generalization, it is not clear what an intermediate result of a general synthesis procedure would be. Future work on a step-level approach may further improve our model.

\section{Conclusion}
\label{sec:conclusion}

We presented a learning-based method to enable approximate but scalable network configuration synthesis. For BGP/OSPF routing, our neural synthesizer is up to $490\times$ faster than SMT-based methods, while producing configurations with very high specification consistency.
We believe there are future research that can be explored in the direction of learning-based synthesis and ML-guided network configuration. \textbf{Ethical Issues} This work does not raise any ethical issues.

\textbf{Acknowledgments} This work was partially supported by an ETH Research Grant ETH-03 19-2.



\newpage
\bibliographystyle{ACM-Reference-Format}
\bibliography{refs}

\section*{Checklist}

The checklist follows the references.  Please
read the checklist guidelines carefully for information on how to answer these
questions.  For each question, change the default \answerTODO{} to \answerYes{},
\answerNo{}, or \answerNA{}.  You are strongly encouraged to include a {\bf
justification to your answer}, either by referencing the appropriate section of
your paper or providing a brief inline description.  For example:
\begin{itemize}
  \item Did you include the license to the code and datasets? \answerYes We plan to publish our code and dataset under an open source license and make it available to others.
\end{itemize}
Please do not modify the questions and only use the provided macros for your
answers.  Note that the Checklist section does not count towards the page
limit.  In your paper, please delete this instructions block and only keep the
Checklist section heading above along with the questions/answers below.

\begin{enumerate}

\item For all authors...
\begin{enumerate}
  \item Do the main claims made in the abstract and introduction accurately reflect the paper's contributions and scope?
    \answerYes{}
  \item Did you describe the limitations of your work?
  \answerYes{}
  \item Did you discuss any potential negative societal impacts of your work?
    \answerNA{}
  \item Have you read the ethics review guidelines and ensured that your paper conforms to them?
    \answerYes{}
\end{enumerate}

\item If you are including theoretical results...
\begin{enumerate}
  \item Did you state the full set of assumptions of all theoretical results?
  \answerNA{}
        \item Did you include complete proofs of all theoretical results?
        \answerNA{}
\end{enumerate}

\item If you ran experiments...
\begin{enumerate}
  \item Did you include the code, data, and instructions needed to reproduce the main experimental results (either in the supplemental material or as a URL)?
    \answerYes{}
  \item Did you specify all the training details (e.g., data splits, hyperparameters, how they were chosen)?
    \answerYes{}
        \item Did you report error bars (e.g., with respect to the random seed after running experiments multiple times)?
    \answerYes{}
        \item Did you include the total amount of compute and the type of resources used (e.g., type of GPUs, internal cluster, or cloud provider)?
    \answerYes{}
\end{enumerate}

\item If you are using existing assets (e.g., code, data, models) or curating/releasing new assets...
\begin{enumerate}
  \item If your work uses existing assets, did you cite the creators?
    \answerNA{}
  \item Did you mention the license of the assets?
  \answerNA{}
  \item Did you include any new assets either in the supplemental material or as a URL?
    \answerYes{}
  \item Did you discuss whether and how consent was obtained from people whose data you're using/curating?
  \answerNA{}
  \item Did you discuss whether the data you are using/curating contains personally identifiable information or offensive content?
  \answerNA{}
\end{enumerate}

\item If you used crowdsourcing or conducted research with human subjects...
\begin{enumerate}
  \item Did you include the full text of instructions given to participants and screenshots, if applicable?
  \answerNA{}
  \item Did you describe any potential participant risks, with links to Institutional Review Board (IRB) approvals, if applicable?
  \answerNA{}
  \item Did you include the estimated hourly wage paid to participants and the total amount spent on participant compensation?
  \answerNA{}
\end{enumerate}

\end{enumerate}

\newpage
\appendix
\section*{Appendix}

\section{Implementation and Model Details}

\subsection{Structural Fact Base Embedding}
\label{app:embedding-scheme}

\begin{figure}[b]
    \footnotesize
    \def\arraystretch{1.2}
    \hspace{-0.45cm}\begin{tabular}{lr}
        $emb(f) := \mathbf{V_f}$ & \textit{(learned fact type embedding)}\\
        $emb(b) := \mathbf{W_{\text{bool}}}\;onehot(b)$ & \textit{(learned boolean embedding)}\\
        $emb(f,i,v) := \mathbf{W_{f_i}}\;onehot(v)$ & \textit{(learned integer embedding)}\\
        $emb(f,i,?) := \mathbf{V_{hole}}$ & \textit{(learned hole embedding)}\vspace{2pt}\\
        $\mathlarger{
        h_c := \sum_{f(c) \in \mathcal{F}}
        emb(f)
        }$ & \textit{(constant embedding)}\vspace{2pt}\\
        $h_{f(a_0, \dots, a_n)} := emb(f) + emb([f(a_0, \dots, a_n)]_{\mathcal{B}}) + \hspace{-1.0cm}\mathlarger{\sum_{(i,v) \in \mathcal{I}(f(a_0, \dots, a_n))} \hspace{-0.95cm}emb(f, i, v) 
        }$\hspace{-2cm} & \textit{(fact embedding)}\\
        $\mathcal{I}(f(a_0, \dots, a_n)) := \{(i,v) \; | \; f(a_0, \dots, a_{i-1}, v, \dots, a_n) \in \mathcal{F} \wedge v:Int \wedge i \in \mathbb{N}$ \}& \textit{(integer arguments)}\vspace{2pt}\\
        $N_i(h_c) := \{h_f \; | \; \exists i \in \mathbb{N} . f(a_0, \dots, a_{i-1}, c, a_{i+1}, \dots, a_n) \in \mathcal{F}\}$& \textit{(constant node neighbors)} \\
        $N_i(h_f) := \{h_c \; | \; \exists i \in \mathbb{N} . f(a_0, \dots, a_{i-1}, c, a_{i+1}, \dots, a_n) \in \mathcal{F}\}$& \textit{(fact node neighbors)} \\
    \end{tabular}
    \caption{Translating a fact base $\mathcal{F}$ to node features. $f$ denotes a fact type, $v$ integer values, $b$ boolean values as 0 or 1, $i$ argument indices and $[f(a)]_{\mathcal{B}}$ the truth value of a fact. }
    \label{fig:graph-embedding}
\end{figure}

To transform a fact base into a graph with node features, we employ a structurally-defined embedding scheme. Given a fact base $\mathcal{F}$, we apply the rules in \cref{fig:graph-embedding} to obtain node features and adjacency information. In the resulting graph, both a fact $f$ and a constant $c$ are represented as distinct nodes with features $h_c$ and $h_f$, respectively. The relationships between facts and constants is encoded as node adjacency as defined by the neighborhood functions $N_i$. We use multiple neighborhood functions, i.e. multiple types of edges, to encode the argument position of a constant occurring in a fact.  Hence, the number of neighborhood functions corresponds to the maximum number of arguments in a single fact, e.g. $3$ in \cref{fig:datalog-encoding}. Integer arguments and unary facts are directly accounted for in the respective node or fact embedding. To handle unknown parameters, we replace the corresponding $emb(...)$ term with the learned $\mathbf{V_{hole}}$ embedding.

Considering the embedding function $emb$, the learned parameters of our embedding are $\mathbf{V_f} \in \mathbb{R}^D$ per fact type in~$\mathcal{F}$, $\mathbf{W_{\text{bool}}} \in \mathbb{R}^{D \times 2}$ for boolean values, $\mathbf{W_{f_i}} \in \mathbb{R}^{D \times N}$ per integer argument of fact type $f$ and $\mathbf{V_{hole}} \in \mathbb{R}^D$ to represent unknown parameters. Hyperparameter $D$ represents the dimension of the latent space the synthesizer model operates in and $N$ specifies the number of supported integer values. 

\subsection{Graph Attention Layer}
\label{app:gat-model-detail}

The recurrence relation of a node $j$'s representation $h_j$ is defined as:

$$
\begin{array}{rcl}
z_j &:=& h_j + Drop(\mathlarger{\sum_{N_i}} h'_{GAT_i})\\
h'_j &:=& FFN(Norm(z_j))\vspace{11pt}\\
FFN(x_j) &:=& Norm(Lin2(Drop(Lin1(x_j)))))
\end{array}
$$

$h'_{GAT_i}$ represents the output of a graph attention layer operating with neighborhood $N_i$ according to \cite{velivckovic2017graph}, $h_j$ the representation of node $j$ at the previous step. $Norm$ refers to batch normalization \cite{ioffe2015batch}, $Drop$ to dropout regularization \cite{hinton2012improving} and $Lin1$ and $Lin2$ to linear layers with an inner dimension of $4N$. The mechanic of combining the results of multiple graph attention layers $h'_{GAT_i}$ directly corresponds to how multiple attention heads are combined in the original Transformer architecture \cite{vaswani2017attention}.

\subsection{Dataset Generation and Training}

\label{app:training}
To train a synthesizer model, we need a large number of fact bases encoding a variety of network topologies, configurations and forwarding specifications. Further, we must provide the model with target values for unknown parameters as a supervision signal. This section discusses how we construct such a dataset for BGP/OSPF synthesis and train a corresponding synthesis model.

\paragraph{Dataset Generation} To obtain network configurations and corresponding forwarding specifications we employ a generative process: We first randomly configure BGP and OSPF parameters for some topology. Then we simulate the OSPF and BGP protocol and compute the resulting forwarding plane. Based on this, we extract a forwarding specification which fixes a random subset of forwarding paths, reachability and isolation properties that are satisfied by the forwarding plane.

\paragraph{Topologies} For training, our dataset is based on random topologies with $16$-$24$ routers. We generate the physical layout of these graphs by triangulation of uniformly sampled points in two-dimensional space.

\paragraph{Simulation} To compute the forwarding plane given an OSPF and BGP configuration, we simulate the protocols. For OSPF, we implement shortest path computation to obtain the forwarding plane. For BGP, we implement the full BGP decision process as shown in \cref{fig:bgp-decision}, except for the MED attribute. In addition to the basic networking model discussed in \cref{sec:synthesis-setting}, we implement support for eBGP and iBGP, fully-meshed BGP session layouts as well as layouts relying on route reflection. For more details of these BGP concepts see \cite{bates2000rr}.



\begin{figure}
    \begin{minipage}[t]{0.45\textwidth}
\begin{lstlisting}[basicstyle=\ttfamily\small]
# topology and configuration
router(c5)
router(c3)
router(c1)
router(c0)
network(c6)
network(c7)

external(c8)
external(c9)
external(c10)
external(c11)
external(c12)
external(c13)

route_reflector(c2)
route_reflector(c4)

connected(c2,c4,?)
connected(c2,c5,?)
connected(c2,c3,?)
connected(c4,c5,?)
connected(c4,c3,?)
connected(c4,c0,?)
connected(c5,c1,?)
connected(c5,c0,?)
connected(c3,c0,?)
connected(c3,c1,?)
connected(c1,c0,?)

ibgp(c2,c4)
ibgp(c2,c5)
ibgp(c2,c0)
ibgp(c2,c1)
ibgp(c4,c3)
ibgp(c4,c4)
\end{lstlisting}
\end{minipage}
\begin{minipage}[t]{0.45\textwidth}
\begin{lstlisting}[basicstyle=\ttfamily\small]

ebgp(c2,c12)
ebgp(c5,c9)
ebgp(c5,c10)
ebgp(c5,c13)
ebgp(c0,c8)
ebgp(c0,c11)

bgp_route(c8,c6,?,?,1,?,1,8)
bgp_route(c9,c6,?,?,0,?,1,9)
bgp_route(c10,c6,?,?,0,?,1,10)
bgp_route(c11,c7,?,?,0,?,1,11)
bgp_route(c12,c7,?,?,0,?,1,12)
bgp_route(c13,c7,?,?,2,?,1,13)

# forwarding specification
fwd(c0,c6,c3)
fwd(c3,c6,c2)
fwd(c2,c6,c4)
not fwd(c3,c6,c0)
not fwd(c0,c6,c11)
not fwd(c11,c6,c0)

not trafficIsolation(c1,c0,c7,c6)
not trafficIsolation(c1,c0,c7,c6)
trafficIsolation(c0,c3,c6,c7)
trafficIsolation(c0,c3,c6,c7)

reachable(c5,c6,c10)
reachable(c1,c6,c4)
not reachable(c0,c6,c5)
reachable(c5,c6,c10)
    \end{lstlisting}
    \end{minipage}
    \caption{An example of a fact base encoding a topology, a sketch of a BGP and OSPF configuration and a forwarding specification.}
    \label{fig:bgp-ospf-fact-base}
    \end{figure} 

\paragraph{Fact Base Encoding} We encode synthesis input using a small set of facts. For an example of a corresponding fact base see \cref{fig:bgp-ospf-fact-base}. We use the \texttt{router}, \texttt{external} and \texttt{network} facts to mark routers, external peers and routing destinations respectively. Physical links between routers are encoded as \texttt{connected(R1,R2,W)} facts where \texttt{W} denotes the OSPF link weight. Further, we rely on \texttt{ibgp}, \texttt{ebgp}, \texttt{route\_reflector} and \texttt{bgp\_route} facts to represent different types of BGP sessions and imported routes. Using the notion of unknown parameters as discussed in \cref{sec:graph-embedding}, we encode configuration parameters, e.g., properties of imported routes or OSPF link weights.

\paragraph{Specification Language} We support a small specification language with three types of forwarding predicates:

\begin{itemize}
    \item \texttt{fwd(R1, Net, R2)} specifies that router \texttt{R1} forwards traffic destined for \texttt{Net} to its neighbor \texttt{R2}.
    \item \texttt{reachable(R1, Net, R2)} specifies that traffic destined for \texttt{Net} that passes through \texttt{R1} also passes through \texttt{R2} before reaching its destination.
    \item \texttt{trafficIsolation(R1, R2, N1, N2)} specifies that only one of $[\text{\texttt{fwd(R1,N1,R2)}}]_{\mathcal{B}}$ and $[\text{\texttt{fwd(R1,N2,R2)}}]_{\mathcal{B}}$ can be true at a time.
\end{itemize}

In the negative case of each predicate, the opposite must hold true for the fact to be satisfied. This specification language can easily be extended by including new specification predicates in the dataset generation process. One must simply provide a method of extracting positive and negative cases of a specification fact from a given forwarding plane. Once included in the training dataset of a synthesizer model, the resulting synthesis system will be able to consider specifications relying on the new type of predicates.

\paragraph{Training} We generate a dataset of $10,240$ input/output samples and instantiate our synthesizer model with a hidden dimension $D = 64$ and supported parameter values $N = 64$. For optimization, we use the Adam optimizer \cite{kingma2014adam} with a learning rate of $10^{-4}$. We stop training after the specification consistency on a validation dataset no longer increases at $\sim2800$ epochs.

\begin{minipage}{1.0\textwidth}
\paragraph{BGP Decision Process}
\label{app:bgp-decision-process}

For completeness, we include the BGP decision process as assumed for our synthesis setting. The shown rules are applied in order until a single best BGP announcements remains.

\begin{enumerate}
    \setlength{\itemsep}{0pt}
    \item Highest Local preference
    \item Lowest AS path length
    \item Prefer Origin (IGP $>$ EBGP $>$ INCOMPLETE)
    \item Lowest Multi-exit discriminator (MED)
    \item External over internal announcements
    \item Lowest IGP cost to egress
    \item Lowest BGP peer id (tie breaker)
\end{enumerate}
\end{minipage}

\newpage
\section{More Evaluation Results}
\label{app:more-eval}

\subsection{Number Of Samples}
\begin{figure}
    \centering
    \begin{tikzpicture}
        \definecolor{color0}{rgb}{0.12156862745098,0.466666666666667,0.705882352941177}
\definecolor{color1}{rgb}{1,0.498039215686275,0.0549019607843137}
\definecolor{color2}{rgb}{0.172549019607843,0.627450980392157,0.172549019607843}
        \begin{axis}[
        height=5cm, width=0.7\columnwidth,
        tick align=outside,
        tick pos=left,
        x grid style={white!69.0196078431373!black},
        xmin=1, xmax=20,
        xtick style={color=black},
        xtick = {1, 5, 10, 15, 20},
        ytick = {0.9, 0.925, 0.95, 0.975, 1.0},
        y grid style={white!69.0196078431373!black},
        ymin=0.85, ymax=1.0,
        ytick style={color=black},
        axis y line*=left,
        axis x line*=bottom,
        xlabel = {\scriptsize Number of Samples},
        ylabel = {\scriptsize{Average Specification Consistency}},
        mark options={fill=white},
        legend cell align={left},
        legend style={
        fill opacity=0.8,
        draw opacity=1,
        text opacity=1,
        at={(0.97,0.03)},
        anchor=south east,
        draw=white!80!black,
        axis y discontinuity=crunch,
        },
        ]
        ]
        \addplot [semithick, color0]
table {%
1 0.924649627326684
2 0.928651932654552
3 0.946390459224601
4 0.955071014780156
5 0.955071014780156
6 0.955071014780156
7 0.955071014780156
8 0.955071014780156
9 0.955071014780156
10 0.961218555763763
11 0.961218555763763
12 0.961218555763763
13 0.961218555763763
14 0.961218555763763
15 0.961218555763763
16 0.961218555763763
17 0.961218555763763
18 0.961218555763763
19 0.961218555763763
20 0.961218555763763
};
\addlegendentry{Small}
\addplot [semithick, color1]
table {%
1 0.951432595182595
2 0.960567210567211
3 0.960567210567211
4 0.971624902874903
5 0.97389763014763
6 0.97389763014763
7 0.97389763014763
8 0.975633741258741
9 0.975633741258741
10 0.975633741258741
11 0.975633741258741
12 0.975633741258741
13 0.975633741258741
14 0.975633741258741
15 0.975633741258741
16 0.975633741258741
17 0.975633741258741
18 0.975633741258741
19 0.975633741258741
20 0.975633741258741
};
\addlegendentry{Medium}
\addplot [semithick, color2]
table {%
1 0.917238477099615
2 0.925898545618838
3 0.925898545618838
4 0.929419672379401
5 0.929419672379401
6 0.929419672379401
7 0.929419672379401
8 0.929419672379401
9 0.931108861568591
10 0.932894575854305
11 0.932894575854305
12 0.932894575854305
13 0.932894575854305
14 0.932894575854305
15 0.932894575854305
16 0.932894575854305
17 0.934910704886563
18 0.934910704886563
19 0.940192395027408
20 0.940192395027408
};
\addlegendentry{Large}
        \end{axis}
    \end{tikzpicture}
    \caption{Average Best Specification Consistency with increasingly many samples from our synthesizer model with $3 \times 16$ BGP/OSPF requirements.}
    \label{fig:eval-num-samples}
\end{figure}
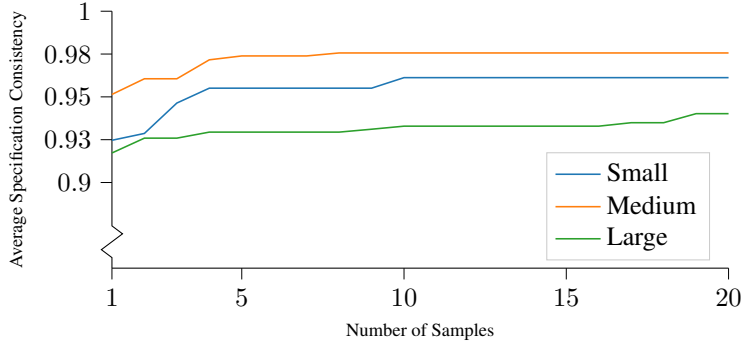

\cref{fig:eval-num-samples} shows the average best consistency across our three datasets with an increasing number of samples for $3\times 16$ BGP/OSPF requirements. As the number of samples increases, the graph indicates the best consistency value reached so far. We observe that sampling more than once improves the average specification consistency across all datasets. With increasing size of the topology, more samples appear to be necessary for the resulting best specification consistency to converge. Therefore, we note that increasing the number of samples can improve consistency but it will also affect overall synthesis time as the model needs to be executed again for each sample. In our other experiments we rely on 4/5 samples per specification, which we consider a good trade-off of fast synthesis time and good specification consistency.

\subsection{Unsatisfiable Specifications} 
\label{app:eval-unsat}
To asses how our model performs in the presence of unsatisfiable specifications, we evaluate its synthesis performance on multiple datasets of unsatisfiable OSPF synthesis tasks.


\paragraph{Specifications} We construct the datasets {\scshape Unsat-$N$}, where $N \in \{2,4,6,8,10,12,14,16\}$, by first generating 16 solvable OSPF forwarding path requirements per topology in Small, Medium and Large. Then we replace $N$ of the requirements per topology with paths obtained under other, random link weights. We repeat this process up to five times, until we can verify that all variants of a topology as contained in the different {\scshape Unsat-$N$} datasets are indeed unsatisfiable using the SMT-based synthesizer NetComplete \cite{el2018netcomplete}. Overall, this leaves us with 8 {\scshape Unsat-$N$} datasets of OSPF synthesis tasks, where we expect the maximum achievable specification consistency to decrease with increasing $N$. For some topologies, we were not successful in generating unsatisfiable variants for all $N$ and therefore we removed them from the considered set of topologies for this part of our evaluation. We count 15 topologies per {\scshape Unsat-$N$} dataset, where the number of nodes ranges between 16 and 153.

\paragraph{Methodology} For synthesis, we use the same sampling configuration as in \cref{sec:synthesis-quality} where we again report the best consistency value across 5 different runs of our synthesis model per synthesis task. We further configure our trained synthesizer model to do OSPF synthesis by only predicting link weights.

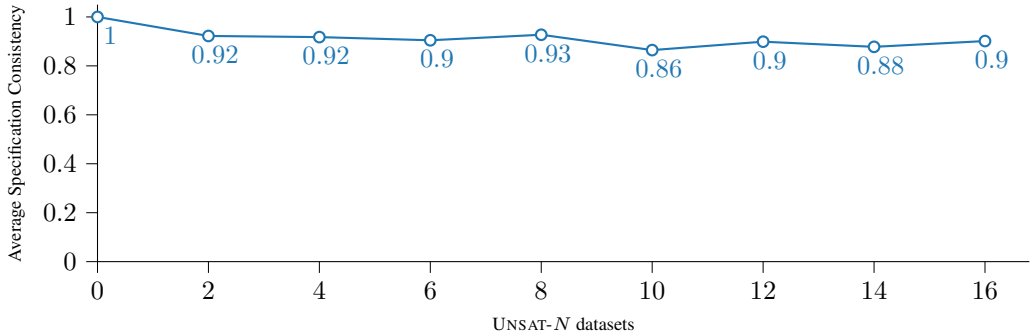
\begin{figure}
    \begin{tikzpicture}
        \definecolor{color0}{rgb}{0.12156862745098,0.466666666666667,0.705882352941177}
        \begin{axis}[
        height=5cm, width=\columnwidth,
        tick align=outside,
        tick pos=left,
        x grid style={white!69.0196078431373!black},
        xmin=0.0, xmax=16.8,
        xtick style={color=black},
        ytick = {0, 0.2, 0.4, 0.6, 0.8, 1.0},
        y grid style={white!69.0196078431373!black},
        ymin=0.0, ymax=1.05,
        ytick style={color=black},
        axis y line*=left,
        axis x line*=bottom,
        xlabel = {\scriptsize {\scshape Unsat-$N$} datasets},
        ylabel = {\scriptsize{Average Specification Consistency}},
        mark=*,
        mark options={fill=white},
        nodes near coords,
        visualization depends on=\thisrow{alignment} \as \alignment,
        every node near coord/.style={anchor=\alignment}
        ]
        ]
        \addplot [thick, color0]
        table {%
        x y alignment
        0 1 125
        2 0.922220307184296 110
        4 0.917586614197893 110
        6 0.904582687887104 110
        8 0.927219906749544 110
        10 0.864502987054147 110
        12 0.898820891594774 110
        14 0.877806848769488 110
        16 0.901392498553868 110
        };
        \end{axis}
    \end{tikzpicture}
    
    \caption{Average OSPF specification consistency of our synthesizer model when applied to unsatisfiable specifications.}
    \label{fig:eval-unsat}
\end{figure}

\paragraph{Results} We report the results of applying our synthesizer model to unsatisfiable OSPF specifications in \cref{fig:eval-unsat}. For comparison, we also include the results of applying our synthesizer model to an {\scshape Unsat-$0$} dataset, i.e. a dataset of satisfiable OSPF synthesis tasks. The average specification consistency drops with unsatisfiable specifications which is expected since the theoretical upper bound for specification consistency is lower with conflicting requirements. At the same time, specification consistency remains high, which suggests that our model still attempts to maximize specification consistency, even when provided with an unsatisfiable specification with many conflicting requirements. In contrast, SMT-based synthesizers like NetComplete will not be able to produce any configurations for the synthesis tasks in our {\scshape Unsat-$N$} datasets.

\begin{minipage}[t]{\textwidth}

\subsection{BGP/OSPF Synthesis Time (GPU)}

In the following table we report BGP/OSPF consistency and synthesis time of our method (Neural) running on a GPU, compared with the SMT-based state-of-art synthesis tool NetComplete. The notation {\tiny{n/8 TO}} indicates the number of timed out synthesis runs out of 8 (25+ minutes). Dataset Large is omitted due to GPU memory restrictions.

\vspace{7pt}
\label{app:bgp-gpu-performance}
\begin{center}
    \textbf{\added{BGP Synthesis Time (GPU, 5 samples)}}\\\footnotesize
    \makebox[\linewidth]{
    \begin{tabular}{rrlllllllll}
        \toprule
        &&\multicolumn{3}{c}{Synthesis Time (s)} & \multicolumn{2}{c}{Accuracy (Neural)} \\
        \multicolumn{2}{l}{\# Reqs.} & NetComplete & Neural (GPU) & Speedup &$\emptyset$ Consistency& $\emptyset$ No. Full Matches\\
        \midrule
        2 reqs. & S & 18.07$s\pm$14.55 & 0.44s$\pm$0.26 & \textbf{41.1x} & 1.00$\pm$0.00 & 8/8\\
        & M & 60.86$s\pm$33.39 & 0.74s$\pm$0.69 & \textbf{81.8x} & 0.94$\pm$0.13 & 6/8\\
        \midrule
        8 reqs. & S & 247.69$s\pm$436.90 & 0.91s$\pm$0.70 & \textbf{270.7x} & 0.95$\pm$0.08 & 5/8\\
        & M & >25m\tiny{  8/8 TO} & 1.23s$\pm$0.72 & \textbf{1220.6x} & 0.97$\pm$0.04 & 4/8\\
        \midrule
        16 reqs. & S & 1416.83$s\pm$235.25\tiny{  7/8 TO} & 1.55s$\pm$0.51 & \textbf{913.3x} & 0.92$\pm$0.06 & 1/8\\
        & M & >25m\tiny{  8/8 TO} & 1.65s$\pm$0.51 & \textbf{906.5x} & 0.93$\pm$0.05 & 1/8\\
    \end{tabular}\vspace{5pt}
    }\\
    \end{center}
\end{minipage}

\newpage
\subsection{OSPF Synthesis Time}
\label{app:evaluation-compare-ospf}

We also compare consistency and synthesis time of our method (Neural) with the SMT-based synthesis tool NetComplete doing OSPF-only synthesis. The notation {\tiny{n/8 TO}} indicates the number of timed out synthesis runs out of 8 (25+ minutes). For this experiment, our neural synthesizer can run on the GPU, as OSPF-only synthesis consumes less GPU memory.

\added{
Below, we report results of OSPF-only synthesis sampling from the synthesizer model 4 times and 5 times respectively. We observe that sampling 4 times can be enough with smaller specifications to obtain full matches for all synthesis tasks. For larger specifications and networks, sampling more than 4 times can improve average consistency. Note however, that sampling 5 times for small specifications and topologies can lead to synthesis times that are longer than with NetComplete (0.7x with 2 reqs., 5 samples, dataset S).
}

\begin{center}
    \footnotesize
    \textbf{OSPF Synthesis Time (4 samples)}\\
    \footnotesize
    \makebox[\linewidth]{
    \begin{tabular}{rrlllllllll}
    \toprule
    &&\multicolumn{3}{c}{Synthesis Time (s)} & \multicolumn{2}{c}{Accuracy (Neural)} \\
    \multicolumn{2}{l}{} & NetComplete & Neural (GPU) & Speedup &$\emptyset$ Consistency& $\emptyset$ No. Full Matches\\
    \midrule
2 reqs. & S & 0.27$s\pm$0.09 & 0.09s$\pm$0.00 & \textbf{2.9x} & 1.00$\pm$0.00 & 8/8\\
 & M & 0.40$s\pm$0.09 & 0.10s$\pm$0.00 & \textbf{4.1x} & 1.00$\pm$0.00 & 8/8\\
 & L & 0.94$s\pm$0.33 & 0.14s$\pm$0.08 & \textbf{6.7x} & 1.00$\pm$0.00 & 8/8\\
\midrule
8 reqs. & S & 0.88$s\pm$0.18 & 0.16s$\pm$0.12 & \textbf{5.3x} & 0.98$\pm$0.04 & 6/8\\
 & M & 1.52$s\pm$0.40 & 0.10s$\pm$0.01 & \textbf{14.7x} & 1.00$\pm$0.00 & 8/8\\
 & L & 3.50$s\pm$1.11 & 0.32s$\pm$0.15 & \textbf{11.1x} & 0.98$\pm$0.03 & 4/8\\
\midrule
16 reqs. & S & 1.65$s\pm$0.37 & 0.21s$\pm$0.14 & \textbf{7.9x} & 0.99$\pm$0.01 & 5/8\\
 & M & 2.65$s\pm$0.70 & 0.25s$\pm$0.16 & \textbf{10.8x} & 0.99$\pm$0.01 & 4/8\\
 & L & 566.64$s\pm$772.90\tiny{  3/8 TO} & 0.23s$\pm$0.11 & \textbf{2430.4x} & 0.99$\pm$0.04 & 7/8\\
\end{tabular}
}
\end{center}

\begin{center}
    \footnotesize
    \added{\textbf{OSPF Synthesis Time (5 samples)}}\\
    \footnotesize
    \makebox[\linewidth]{
    \begin{tabular}{rrlllllllll}
    \toprule
    &&\multicolumn{3}{c}{Synthesis Time (s)} & \multicolumn{2}{c}{Accuracy (Neural)} \\
    \multicolumn{2}{l}{} & NetComplete & Neural (GPU) & Speedup &$\emptyset$ Consistency& $\emptyset$ No. Full Matches\\
    \midrule
2 reqs. & S & 0.27$s\pm$0.09 & 0.36s$\pm$0.01 & \textbf{0.7x} & 1.00$\pm$0.00 & 8/8\\
 & M & 0.40$s\pm$0.09 & 0.37s$\pm$0.00 & \textbf{1.1x} & 1.00$\pm$0.00 & 8/8\\
 & L & 0.94$s\pm$0.33 & 0.52s$\pm$0.21 & \textbf{1.8x} & 1.00$\pm$0.00 & 8/8\\
\midrule
8 reqs. & S & 0.88$s\pm$0.18 & 0.70s$\pm$0.63 & \textbf{1.3x} & 0.98$\pm$0.04 & 6/8\\
 & M & 1.52$s\pm$0.40 & 0.37s$\pm$0.00 & \textbf{4.1x} & 1.00$\pm$0.00 & 8/8\\
 & L & 3.50$s\pm$1.11 & 1.69s$\pm$1.08 & \textbf{2.1x} & 0.98$\pm$0.02 & 4/8\\
\midrule
16 reqs. & S & 1.65$s\pm$0.37 & 0.81s$\pm$0.69 & \textbf{2.0x} & 0.99$\pm$0.02 & 6/8\\
 & M & 2.65$s\pm$0.70 & 1.01s$\pm$0.73 & \textbf{2.6x} & 0.99$\pm$0.01 & 5/8\\
 & L & 566.64$s\pm$772.90\tiny{  3/8 TO} & 1.09s$\pm$0.89 & \textbf{518.1x} & 0.99$\pm$0.04 & 6/8\\
\end{tabular}
}
\end{center}

\section{Future Research Directions}    
\label{app:future}
The key challenge with our learning-based system remains its imprecision, i.e. that it produces only partially-consistent configurations in some cases. On one hand, parts of the configuration synthesis problem are NP-hard \cite{bley2007inapproximability,fortz2000internet, vissicchio2012ibgp}, which means that practical and scalable synthesis tools have to compromise on precision unless P=NP. At the same time, satisfying the complete specification may be crucial, depending on the scenario. Therefore, our learning-based synthesis will not be applicable in the same way as precise, SMT-based synthesizers. Based on this insight, we envision a whole range of future directions for which imprecise, learning-based synthesis will be very useful:

\paragraph{ML-assisted Configuration} Imprecise synthesis may be used to enable ML-assisted configuration in the form of a semi-auto\-mated process: users provide a specification, apply the synthesizer and perform additional tweaking after they obtain a sufficiently consistent configuration. Optionally, our synthesizer model can also be applied again to predict alternative values for some of the configuration parameters, by masking only the desired parameters in a configuration (cf. \cref{sec:graph-embedding}). Recently, similar systems for code completion, like GitHub CoPilot \cite{chen2021evaluating, githubCopilotWeb}, have demonstrated that this query-predict-modify workflow can be highly effective at combining an imprecise ML-guided recommendation system with user interaction \cite{drori2021solving}.

\paragraph{Hybrid Systems} Given that further tweaking of a configuration may be necessary, future work may also explore seeding SMT-based, partial synthesis methods \cite{el2018netcomplete} or configuration repair methods \cite{gember2017automatically} with generated candidate solutions. This can further automate the process of obtaining fully consistent configurations while maintaining some of the performance benefits of a learning-based system. Existing work on partial synthesis has already shown that SMT-based synthesis can be much faster when a rough sketch of a configuration is already provided as an input \cite{el2018netcomplete}.

\paragraph{Imprecision in SMT-based synthesis} Tweaking the result of a synthesizer is undesirable, but it is also a common practice with existing SMT-based synthesis, as it can be imprecise too. This can be caused by hand-coded SMT synthesis rules which do not always hold up in practice, given the complex behavior of real-world routers \cite{birkner2021metha}. Similarly, our learning-based method is imprecise, but could actually be trained on real-world data, possibly bridging the gap between an assumed formal model and real-world behavior. Future work is necessary to clarify how the configuration tweaking process differs in practice between SMT-based and learning-based synthesis.

\paragraph{Optimality with Unsatisfiable Specifications} As demonstrated in \cref{app:eval-unsat}, our learning-based system produces highly consistent configurations even in the presence of an unsatisfiable specification. In practice, this can be helpful as operators do not know whether their specification is unsatisfiable ahead of time. In these cases, a partially consistent configuration is the best one can hope for. In contrast, using SMT solvers in this scenario will not produce a configuration at all, while unsatisfiable specifications can be quite difficult to debug.
    
\paragraph{Service Level Agreements} Lastly, the notion of relaxed specification consistency is not entirely unbeknownst to the world of networking. Networks often operate in accordance with so-called Service Level Agreements (SLA), contracts that specify the guaranteed properties of service in terms of relative availability over time. Given this concept of SLAs, partially violated specifications in some scenarios can be tolerable, as explored by previous work like \cite{steffen2020probabilistic}.
\newpage

\section{Additional Experiments}
\label{app:additional-experiments}

In response to the reviews we carried out a number of additional experiments. We will integrate these results into our draft where applicable.

\subsection{Ablation and Parameter Study}

We conducted an ablation study to explore both the effectiveness of our architecture as well as the use of different components (e.g. different GNN layer modules). We compare across the following configurations. Configurations marked with (*) correspond to the configuration presented in the main body of the paper.

\paragraph{Noise and Edge Types} {\scshape NoNoise} corresponds to our model without adding gaussian noise before applying the processor network. {\scshape NoEdgeTypes} corresponds to our model with only one edge type. {\scshape NoiseEdgeTy (*)} correspond to our model as presented in the paper.

\paragraph{GNN Modules} We also experiment with different GNN modules used internally by the encoder and processor network (cf. \cref{sec:graph-transformer}). In addition to a graph attention module (GAT, \cite{velivckovic2017graph}), we also evaluate models that employ a simple message-passing layer (MPNN-max, \cite{gilmer2017neural}) and a graph convolutional layer (GCN-max, \cite{kipf2016semi}), both aggregating by maximization.

\paragraph{Hidden Dimensionality} We also experiment with different values ($16$, $32$, $64$ and $128$) for dimensionality $D$ of the synthesizer model.

\paragraph{Separate Latent Spaces} According to our graph-based encoding, we embed information on topology, configurations and specification all in a common latent space. However, we also experiment with embedding the topology and configuration information into three separate latent spaces. For this, we train models with three different pairs of encoder+processor networks, one per type of facts/nodes relating to topology, configuration and specification respectively. At each layer and synthesizer iteration, the different encoder/processor GNNs can attend to the intermediate representations of the adjacent nodes according to the underlying fact base graph. As this induces a threefold increase in parameters, we compare the results of such synthesizer models {\scshape SepSpace-16}, {\scshape SepSpace-64} and our model {\scshape ComSpace-64} (*). Due to the long training time of these models, we compare after training these configuration for 2000 epochs only. 

\paragraph{Sampling, Dataset and Metrics} For the examined configurations, we train a synthesizer model using the same training setup as described in \ref{app:training}. Using the resulting models, we perform synthesis for all datasets S, M, and L, relying on 5 samples per task. For {\scshape NoNoise} and {\scshape NoiseEdgeTy (*)} we additionally perform synthesis for the same datasets, but with 20 samples. Here, the test datasets S, M and L include all generated synthesis tasks of the specification size classes 3x2, 3x8 and 3x16 as used in other parts of our evaluation. Overall, this leads to 24 synthesis tasks on 8 different topologies per dataset S/M/L. We measure average best specification consistency, as well as number of full (Full) and partial, good ($>90\%$) matches in terms of specification consistency.

\begin{table}
\begin{center}
\textbf{Ablation Results (5 samples)}
\footnotesize
\makebox[\linewidth]{
\begin{tabular}{rlllllllll}
    \toprule
    Configuration & S & Full & $>90\%$ & M & Full & $>90\%$ & L & Full & $>90\%$ \\
    \midrule
    
{\scshape NoiseEdgeTy (*)} & \textbf{0.97$\pm$0.05} & \textbf{15/24} & \textbf{20/24} & \textbf{0.96$\pm$0.05} & \textbf{12/24} & \textbf{19/24} & 0.95$\pm$0.03 & 7/24 & 18/24 \\

{\scshape NoNoise}& 0.96$\pm$0.07 & \textbf{15/24} & 19/24 & 0.95$\pm$0.05 & 10/24 & \textbf{19/24} & \textbf{0.96$\pm$0.04} & \textbf{8/24} & \textbf{21/24} \\

{\scshape NoEdgeTypes} & 0.94$\pm$0.07 & 11/24 & 18/24 & 0.93$\pm$0.08 & 9/24 & 17/24 & ¸0.92$\pm$0.07 & 6/24 & 16/24 \\
\midrule

GAT \cite{velivckovic2017graph} (*)& \textbf{0.96$\pm$0.05} & 12/24 & \textbf{21/24} & \textbf{0.96$\pm$0.06} & \textbf{12/24} & \textbf{20/24} & \textbf{0.94$\pm$0.05} & \textbf{6/24} & \textbf{18/24} \\

MPNN-Max \cite{gilmer2017neural}& \textbf{0.96$\pm$0.05} & \textbf{14/24} & 20/24 & 0.95$\pm$0.05 & 9/24 & 18/24 & 0.93$\pm$0.05 & 5/24 & 14/24 \\

GCN-Max \cite{kipf2016semi} & 0.93$\pm$0.09 & 11/24 & 16/24 & 0.94$\pm$0.08 & 9/24 & 17/24 & 0.92$\pm$0.06 & 4/24 & 14/24 \\
\midrule

Hidden Dim 16& 0.96$\pm$0.05 & 14/24 & 19/24 & 0.93$\pm$0.07 & 7/24 & 18/24 & 0.90$\pm$0.08 & 4/24 & 11/24 \\

Hidden Dim 32& 0.96$\pm$0.05 & 12/24 & 18/24 & 0.94$\pm$0.06 & 8/24 & 18/24 & 0.94$\pm$0.07 & 6/24 & 18/24 \\

Hidden Dim 64 (*)& \textbf{0.97$\pm$0.05} & \textbf{15/24} & 20/24 & \textbf{0.96$\pm$0.05} & \textbf{12/24} & 19/24 & \textbf{0.95$\pm$0.03} & 7/24 & 18/24 \\

Hidden Dim 128& \textbf{0.97$\pm$0.05} & \textbf{15/24} & 19/24 & \textbf{0.96$\pm$0.05} & \textbf{12/24} & \textbf{20/24} & \textbf{0.95$\pm$0.06} & \textbf{9/24} & \textbf{20/24} \\
\midrule

{\scshape ComSpace-64 (*)} & \textbf{0.96$\pm$0.05} & 14/24 & \textbf{20/24} & \textbf{0.95$\pm$0.06} & \textbf{11/24} & 19/24 & 0.95$\pm$0.05 & 8/24 & 19/24 \\

{\scshape SepSpace-64} & 0.95$\pm$0.09 & \textbf{15/24} & \textbf{20/24} & \textbf{0.95$\pm$0.06} & 10/24 & 18/24 & \textbf{0.96$\pm$0.03} & \textbf{9/24} & \textbf{21/24} \\

{\scshape SepSpace-16} & 0.94$\pm$0.07 & 10/24 & 16/24 & \textbf{0.95$\pm$0.05} & \textbf{11/24} & \textbf{21/24} & 0.93$\pm$0.07 & 6/24 & 17/24 \\
\midrule

\midrule

\end{tabular}\vspace{5pt}
}
\end{center}
\caption{The main results of our ablation and parameter study.}
\label{fig:ablation-main}
\end{table}

\begin{table}
\begin{center}
    \textbf{Ablation Results (20 samples)}
    \footnotesize
    \makebox[\linewidth]{
    \begin{tabular}{llllllllll}
        \toprule
        Configuration & S & Full & $>90\%$ & M & Full & $>90\%$ & L & Full & $>90\%$ \\
        \midrule
    {\scshape NoiseEdgeTy (*)} & \textbf{0.98$\pm$0.03} & 16/24 & \textbf{22/24} & \textbf{0.99$\pm$0.03} & \textbf{16/24} & \textbf{22/24} & \textbf{0.97$\pm$0.03} & \textbf{12/24} & \textbf{23/24} \\
    \midrule

    {\scshape NoNoise} & 0.97$\pm$0.06 & \textbf{18/24} & \textbf{22/24} & 0.97$\pm$0.04 & 12/24 & 21/24 & 0.96$\pm$0.05 & 9/24 & 19/24 \\

    \end{tabular}\vspace{5pt}
    }
\end{center}
\caption{Synthesis performance when sampling 20 times from a synthesizer model with and without adding noise ({\scshape NoiseEdgeTy (*)} vs. {\scshape NoNoise}).}
\label{fig:ablation-20-samples}
\end{table}

\paragraph{Results} The results of our experiments are provided in \cref{fig:ablation-main} and \cref{fig:ablation-20-samples}. For supporting multiple edge types in {\scshape NoiseEdgeTy}, we observe a clear benefit over {\scshape NoEdgeTypes}, both with respect to average best consistency as well as the total number of full and good matches. With just 5 samples, the effect of {\scshape NoNoise} is not very pronounced. However when increasing the number times we sample from the synthesizer model per task, we observe a clear performance improvement. When sampling up to 20 times from the model, {\scshape NoiseEdgeTy (*)} outperforms {\scshape NoNoise} in almost all metrics (cf. \cref{fig:ablation-20-samples}), leading to higher average consistency and a higher number of full and good matches. We hypothesize that adding noise helps the model in producing a wider variety of solutions, eventually leading to better results when selecting the best one.

Regarding the choice of GNN module, we observe that GAT is the most effective, but MPNN-max can also be a good choice. 

Regarding the choice of hidden dimension, we observe that 64 is a good balance of performance and memory usage, while 128 only brings slight improvements. For our purposes we selected $D=64$, since this can improve synthesis time significantly when running on a GPU (cf. \cref{app:bgp-gpu-performance}). For $D=128$ this may not always be possible, especially when limited to a single GPU and working with large topologies.

Lastly, we cannot observe a significant benefit of separating the latent spaces of topology, configuration and specification as described above using multiple encoder+processer networks. {\scshape SepSpace-64} appears to perform mostly on par with our {\scshape ComSpace-64} configuration, while having significantly more parameters. {\scshape SepSpace-16} is more comparable in terms of the number of parameters, but it performs worse than {\scshape ComSpace-64} in most metrics.

\subsection{Dataset Statistics and Distribution Shift}

To examine the synthesis performance of our model with real topologies, we source our evaluation datasets from the Topology Zoo \cite{knight2011internet}, a collection of real-world topologies. For training on the other hand, we only rely on synthetic data based on random topologies, configurations and specifications that are easy to generate by simulation (cf. \ref{app:training}). As a consequence, we observe distribution shift between the synthetic data we train on and the closer-to-real-world data we evaluate~on. 

\begin{figure}[tb]
    \centering
    \includegraphics[width=0.6\textwidth]{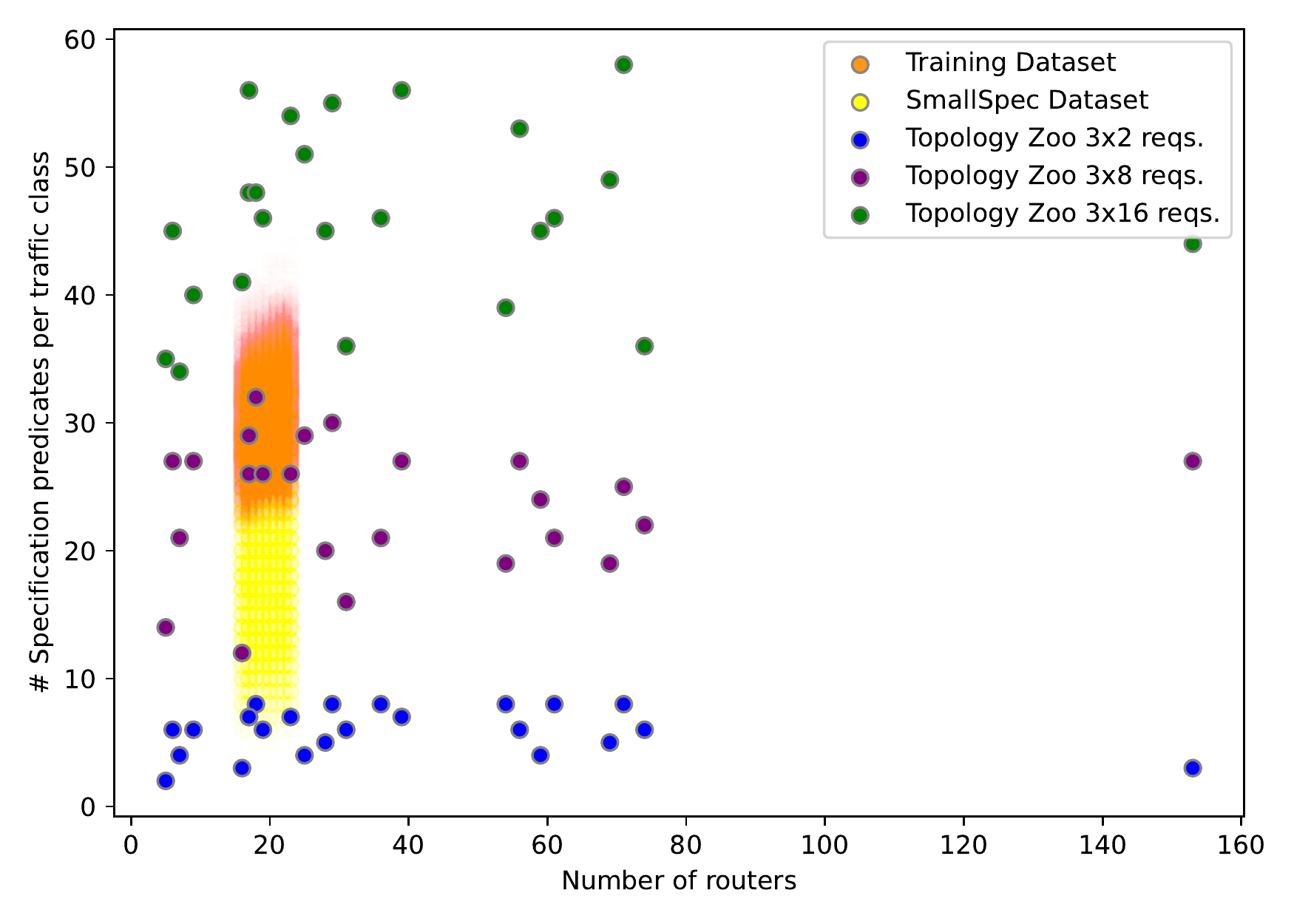}
    \caption{Dataset distribution of network and specification size in terms of routers and the number of specification predicates per traffic class respectively.}
    \label{fig:dataset-stats}
\end{figure}

\paragraph{Training and Evaluation Datasets} \cref{fig:dataset-stats} illustrates parts of the distribution shift with respect to the number of nodes and specifications size. Our synthetic training dataset (Training Dataset) only contains samples of very limited size (15-25 nodes) with $\sim$20-40 different specification predicates per traffic class. In contrast, our Topology Zoo datasets used for evaluation contain topologies of much larger size (5-153) and larger specifications ($\sim$1-60 predicates per traffic class). Considering this distribution shift and the good performance of learned synthesis on the evaluation datasets, our model appears to generalize well to larger topologies/specifications, even without having seen similarly-sized problem instances during training. Comparably strong generalization properties have also been previously observed with other NAR-based models \cite{velivckovic2019neural}. 

\paragraph{Specification Distribution Shift (SmallSpec)} Our evaluation datasets rely on real topologies but have to resort to synthetic specifications due to the lack of a large, practical dataset in this space. Nonetheless, we want to provide some preliminary insight into the possible effect of a specification distribution shift, as would be the case when applying our model to real-world synthesis tasks. For this, we construct a SmallSpec dataset (cf. \cref{fig:dataset-stats}) which is similar to our training dataset but with smaller specifications. In \cref{fig:smallspec-distribution-shift} we report the results of training a synthesizer model on the SmallSpec dataset and comparing its performance with a model trained on our regular Training Dataset. Both models are evaluated on the same Topology Zoo evaluation datasets as in our evaluation. The resulting SmallSpec model achieves slightly lower but comparable synthesis performance. This provides some evidence regarding the robustness of our models when it comes to a distribution shift of the specifications used for synthesis.

\begin{table}
\begin{center}
    \textbf{SmallSpec Distribution Shift in Training (5 samples)}
    \footnotesize
    \makebox[\linewidth]{
    \begin{tabular}{rlllllllll}
        \toprule
        Configuration & S & Full & $>90\%$ & M & Full & $>90\%$ & L & Full & $>90\%$ \\
        \midrule
        
    Training Dataset (*)& \textbf{0.97$\pm$0.05} & \textbf{15/24} & 20/24 & \textbf{0.96$\pm$0.05} & \textbf{12/24} & \textbf{19/24} & \textbf{0.95$\pm$0.03} & \textbf{7/24} & 18/24 \\
    
    SmallSpec & \textbf{0.97$\pm$0.04}&14/24& \textbf{22/24} & 0.95$\pm$0.06&10/24&18/24 & 0.94$\pm$0.05&6/24&\textbf{19/24} \\
    \midrule
    
    \end{tabular}\vspace{5pt}
    }
    \end{center}
    \caption{Examining the effect of specification distribution shift by comparing synthesis performance of a model trained on the SmallSpec dataset (smaller specifications) with a model trained on our regular training dataset.}
    \label{fig:smallspec-distribution-shift}
\end{table}

\subsection{Varying Number of Samples}

In a previous revision of the paper, our evaluation relied on 5 samples to obtain consistency results and only 4 to determine synthesis times. In practice, this can be a sensible choice, trading off accuracy for speed. However, to rectify this discrepancy in our results, we carried out the respective dual experiments to also determine average synthesis times and consistency results with 5 and 4 samples respectively. The experiments in the main body of this revision all rely on 5 samples.

\paragraph{Consistency} We compare synthesis performance when sampling 4/5 times from our synthesis model. Below we list the results for average best consistency, number of full matches as well as the number of $>90\%$ matches for each of the datasets S, M, L. The results correspond directly to the results in column "Overall" in \cref*{fig:eval-consistency}, where we report the results for sampling 5 times.

\begin{center}
\textbf{BGP/OSPF Consistency (4 samples vs. 5 samples)}
\footnotesize
\makebox[\linewidth]{
\begin{tabular}{rrlllllllll}
    \toprule
    Req. & Samples & S & Full & $>90\%$ & M & Full & $>90\%$ & L & Full & $>90\%$ \\
    \midrule
3x2 & 4 samples & \textbf{0.96$\pm$0.07} & \textbf{6/8} & \textbf{6/8} & 0.91$\pm$0.11 & 4/8 & 4/8 & 0.91$\pm$0.11 & \textbf{5/8} & \textbf{5/8} \\

         & 5 samples & \textbf{0.96$\pm$0.07} & \textbf{6/8} & \textbf{6/8} & \textbf{0.94$\pm$0.08} & \textbf{5/8} & \textbf{5/8} & \textbf{0.94$\pm$0.006} & 4/8 & 4/8 \\

        \midrule
3x8 & 4 samples & \textbf{0.97$\pm$0.04} & 3/8 & \textbf{7/8} & \textbf{0.98$\pm$0.02} & 3/8 & \textbf{8/8} & \textbf{0.95$\pm$0.04} & \textbf{1/8} & 7/8 \\

         & 5 samples & 0.96$\pm$0.04 & \textbf{4/8} & \textbf{7/8} & \textbf{0.98$\pm$0.03} & \textbf{4/8} & \textbf{8/8} & \textbf{0.95$\pm$0.03} & \textbf{1/8} & \textbf{8/8} \\
        
         \midrule
3x16 & 4 samples & \textbf{0.95$\pm$0.03} & \textbf{2/8} & \textbf{8/8} & 0.95$\pm$0.04 & 2/8 & 6/8 & \textbf{0.95$\pm$0.04} & \textbf{1/8} & \textbf{6/8} \\
    
         & 5 samples & \textbf{0.95$\pm$0.03} & \textbf{2/8} & \textbf{8/8} & \textbf{0.96$\pm$0.04} & \textbf{3/8} & \textbf{7/8} & 0.93$\pm$0.05 & \textbf{1/8} & \textbf{6/8} \\
\end{tabular}\vspace{5pt}
}
\end{center}

These results suggest that a higher number of samples can increase average consistency as well as the number of full and good, partial matches. Especially small specifications (cf. 3x2), seem to benefit from more samples. Overall, however there is not a very large difference between sampling 4/5 times. 

\paragraph{Synthesis Time} We also compare synthesis time of sampling 4/5 times below. We report the synthesis time of running our synthesizer model with 4 samples as reported in the paper. Next, we report synthesis time of running our synthesizer model with 5 samples. Last, we report the speedup over NetComplete when running our synthesizer model with 4 and 5 samples, respectively.

\begin{center}
    \textbf{BGP/OSPF Synthesis Time (4 samples vs. 5 samples)}
    \footnotesize
    \makebox[\linewidth]{
    \begin{tabular}{rrlllllllll}
    \toprule
    \multicolumn{2}{l}{\# Requirements} & 4 samples (s) & 5 samples (s) & Speedup (4 samples) & Speedup (5 samples)\\
    \midrule
    2 reqs. & S & 0.64s$\pm$0.38 & 0.72s$\pm$0.54 & \textbf{28.2x} & 25.2x\\
    & M & 2.75s$\pm$3.29 & 3.18s$\pm$4.32 & \textbf{22.2x} & 19.1x\\
    & L & 22.30s$\pm$26.86 & 24.25s$\pm$28.35 & \textbf{62.3x} & 57.3x\\
    \midrule
    8 reqs. & S & 1.07s$\pm$0.84 & 1.25s$\pm$1.02 & \textbf{232.5x} & 198.7x\\
    & M &3.47s$\pm$3.34 & 4.55s$\pm$4.30 & \textbf{432.2x} & 329.8x\\
    & L &30.96s$\pm$28.18 & 31.28s$\pm$28.53 & \textbf{48.4x} & 48.0x\\
    \midrule
    16 reqs. & S & 2.53s$\pm$1.85 & 2.88s$\pm$1.66 & \textbf{560.5x} & 492.0x\\
    & M & 5.48s$\pm$3.90 & 6.53s$\pm$5.10 & \textbf{273.7x} & 229.8x\\
    & L & 69.09s$\pm$108.17 & 87.99s$\pm$141.97 & \textbf{21.7x} & 17.0x\\
    \end{tabular}
    }
\end{center}

Overall, running our synthesizer model with 5 samples means that the model is invoked one more time per synthesis task. This is clearly reflected by the resulting synthesis times. However, as the number of samples is only a linear factor for overall synthesis time, the speedup over NetComplete remains very significant.


\end{document}






\appendix
\twocolumn[
\section{BGP Synthesis Time (GPU)}
\vspace{20pt}
\label{sec:bgp-gpu-performance}
\begin{center}
    \begin{minipage}{0.9\textwidth}
        \textbf{BGP Synthesis Time (GPU)}\\
        \footnotesize
        \begin{tabular}{rrccccccccc}
        \toprule
        &&\multicolumn{3}{c}{Synthesis Time (s)} & \multicolumn{2}{c}{Accuracy (Neural)} \\
        \multicolumn{2}{l}{\# Requirements} & NetComplete & Neural (GPU) & Speedup &$\emptyset$ Consistency& $\emptyset$ No. Full Matches\\
            \midrule
        2 requirements &Small & 18.073s & \textbf{0.496s} & 36.4x & 0.969 & 0.875\\
        &Medium & 60.864s & \textbf{0.676s} & 90.1x & 0.939 & 0.75\\
        \midrule
        8 requirements &Small & 247.691s & \textbf{0.719s} & 344.7x & 0.96 & 0.75\\
        &Medium & 1500s\tiny{  8/8 TO} & \textbf{1.053s} & 1424.3x & 0.968 & 0.5\\
        \midrule
        16 requirements &Small & 1416.828s\tiny{  7/8 TO} & \textbf{1.296s} & 1093.0x & 0.922 & 0.125\\
    &Medium & 1500s\tiny{  8/8 TO} & \textbf{1.38s} & 1086.6x & 0.945 & 0.125\\
    \end{tabular}\vspace{5pt}
    {
        \footnotesize
        Comparing consistency and synthesis time of our method (Neural) running on a GPU with the SMT-based state-of-art synthesis tool NetComplete. The notation {\tiny{n/8 TO}} indicates the number of timed out synthesis runs out of 8 (25+ minutes). Dataset Large is omitted due to GPU memory restrictions.
        }
    \end{minipage}
\end{center}
]
